\documentclass[aps,prb,amsmath,amssymb,floatfix,reprint]{revtex4-1}
\usepackage{graphicx}
\usepackage{subfig}
\usepackage{bm}
\usepackage{amsfonts,mathrsfs}
\usepackage{hyperref}
\hypersetup{colorlinks,citecolor=blue,linkcolor=blue,urlcolor=blue}
\newcommand{\ital}{\textit}
\newcommand{\bd}{\textbf}

\begin{document}
\title{On the Accuracy of van der Waals Inclusive Density-Functional Theory Exchange-Correlation Functionals for Ice at Ambient and High Pressures}
\author{Biswajit Santra$^{1,2}$}
\author{Ji\v{r}\'{i} Klime\v{s}$^{3}$}
\author{Alexandre Tkatchenko$^1$}
\author{Dario Alf\`e$^{4,5,6}$}
\author{Ben Slater$^7$}
\author{Angelos Michaelides$^{4,7}$}\email{angelos.michaelides@ucl.ac.uk}
\author{Roberto Car$^{2}$}
\author{Matthias Scheffler$^1$}
\affiliation{
$^1$Fritz-Haber-Institut der Max-Planck-Gesellschaft, Faradayweg 4-6, 14195 Berlin, Germany\\
$^2$Department of Chemistry, Princeton University, Princeton, New Jersey 08544, USA\\
$^3$Faculty of Physics and Center for Computational Materials Science, University of Vienna, Sensengasse 8/12, A-1090 Wien, Austria\\
$^4$London Centre for Nanotechnology, University College London, London WC1E6BT, UK\\
$^5$Department of Physics and Astronomy, University College London, London WC1E6BT, UK\\
$^6$Department of Earth Sciences, University College London, London WC1E6BT, UK\\
$^7$Department of Chemistry, University College London, London WC1E6BT, UK
}

\begin{abstract}

Density-functional theory (DFT) has been widely used to study water and ice for at least 20 years. 
However, the reliability of different DFT exchange-correlation (\emph{xc}) functionals for water remains a matter of considerable debate. 
This is particularly true in light of the recent development of DFT based methods that account for van der Waals (vdW) dispersion forces. 
Here, we report a detailed study with several~\emph{xc} functionals (semi-local, hybrid, and vdW inclusive approaches) on ice I$h$ and
six proton ordered phases of ice. 
Consistent with our previous study [Phys. Rev. Lett.~\textbf{107}, 185701 (2011)] which showed that vdW forces become increasingly
important at high pressures, we find here that all vdW inclusive methods considered 
improve the relative energies and transition pressures of the high-pressure ice phases
compared to those obtained with semi-local or hybrid~\emph{xc} functionals.
However, we also find that significant discrepancies between experiment and the vdW inclusive approaches remain in the cohesive properties 
of the various phases, causing certain phases to be absent from the phase diagram.
Therefore, room for improvement in the description of water at ambient and high pressures remains and 
we suggest that because of the stern test the high pressure ice phases pose they should be used in
future benchmark studies of simulation methods for water.

\end{abstract}

\maketitle

\section{Introduction}

Density-functional theory (DFT) is now widely used to study water and ice in a range of different environments,
including for example bulk water, water at interfaces, and water under confinement. 
Most DFT studies of water have involved the application of 
semi-local generalized gradient approximations (GGA) for the exchange and correlation (\emph{xc}) energy.
Whilst these studies have proved to be very useful in providing insights into the structure and properties of water,
there are persistent question marks over the quantitative accuracy of such~\emph{xc} functionals,
in particular for the treatment of condensed phase water which is held together by hydrogen (H) bonding and van der Waals (vdW) interactions.
Over the years this has prompted a number of benchmark studies 
focused on gas phase water clusters 
\cite{santra_jcp_2007,santra_jcp_2008,santra_jcp_2009,truhlar_hexamer_2008,goddard_dimer_2004,goddard_hexamer_2004,xantheas_1995,kim_jordan_1994,
tschumper_JCPA_2006,shields_kirschner_2008,perdew_JCPB_2005,truhlar_pbe1w_2005,gillan_jcp_12,joel_pbe,Tsuzuki,Novoa,hammond_09,jordan_10},
liquid water,
\cite{grossman_water_1,grossman_water_2,artacho,McGrath_05,McGrath_06,Mcgrath_jpcA_06,tuckerman_jcp_06,tuckerman_jcp_07,Todorova,vandevondele_05,vandevondele_08,parrinello_09,sit_marzari_jcp_05,artacho_05,asthagiri_03,am05_water_09,Kuo_04,sprik_jcp_96,silvestrlli_jcp_99,xantheas_JCP_09,zhang_jctc_2011_2,parrinello_03,car_08,DFT_MD_93,ojo_cpl_2011,gillan_jcp_13}
and crystalline phases of ice. 
\cite{santra_prl_2011,murray_prl_2012,hamada_10,kolb_prb_11,pccp_ice_12,fang_prb_2013,gillan_arxiv_1303.0751,pamuk_prl_2012,car_ice_92,hamann_97,klein_05,slater_jacs_06,de_koning_06,feibelman_08,hermann_prl_08,pan_08,pisani_09,car_09,militzer_10,labat_11})
%
%
While we currently have a relatively clear understanding about the
performance of various ~\emph{xc} functionals for gas phase clusters,
this is far from being established for ice and liquid water.
%
%
This is particularly true in light of recent work which has shown that 
vdW dispersion forces are important for the accurate description of different properties of water. 
\cite{santra_prl_2011,murray_prl_2012,hamada_10,kolb_prb_11,pccp_ice_12,fang_prb_2013,gillan_arxiv_1303.0751,lin_jpcB_09,schmidt_jpcB_09,wang_jcp_2011,mogelhoj_jpcb_2011,zhang_jctc_2011,jonchiere_jcp_2011,yoo_jcp_11,tuckerman_jcp_2012,zhaofeng_thesis_2012}
Understanding the role of vdW forces in water has been greatly helped by the emergence of various approaches
for accounting for vdW forces within the framework of DFT.
\cite{Dion_vdw_04,TS_vdw_09,klimes_10,oavl_prl_2004,silvestrelli_prl_2008,sato_jcp_2009,grimme_vdw_10,vydrov_prl_2009,becke_jcp_2005,perdew_pnas_2012}
In the last few years many of the vdW inclusive DFT~\emph{xc} functionals have been used 
to investigate the effects of vdW on the structural, energetic, and vibrational properties of liquid water.
\cite{lin_jpcB_09,schmidt_jpcB_09,wang_jcp_2011,mogelhoj_jpcb_2011,zhang_jctc_2011,jonchiere_jcp_2011,yoo_jcp_11,tuckerman_jcp_2012,zhaofeng_thesis_2012}
Overall, with vdW inclusive~\emph{xc} functionals there are indeed improvements 
in certain calculated properties of liquid water. 
For example, the first peak in the oxygen-oxygen radial distribution function is generally reduced and
brought into closer agreement with experiment.
However, the accuracy of the computed properties strongly depends on the methods chosen to incorporate vdW 
as well as the technical details of the molecular dynamics simulations.
There is, of course, also the challenge of accounting for quantum nuclear effects, which is rarely done in~\emph{ab initio} studies
of liquid water.~\cite{parrinello_03,car_08}
However, in contrast to liquid water, the various crystalline phases of ice represent a relatively straightforward set of
structures against which DFT methods can be tested.
Indeed there are at present 15 experimentally characterized ice phases with water molecules in a number of distinct arrangements,
H bond networks, and densities.~\cite{ice_book,ice-15,salzmann_rsc_2007,salzmann_pccp_2011}
Many of the ice phases are complicated with disordered water arrangements (so called ``proton disordered'').
However, some phases have relatively simple proton ordered arrangements of water molecules,
and it is these phases that are particularly suitable as benchmarks. 
%
Furthermore, thanks to Whalley's extrapolations of the experimental finite temperature and pressure
phase coexistence lines to zero temperature, for some of these phases there are even estimates of the internal energy
differences,~\cite{whalley_jcp_84} which makes theoretical benchmarks particularly straightforward and mitigates the need for expensive free
energy calculations.

\begin{figure*}
\begin{center}
\includegraphics[width=12cm]{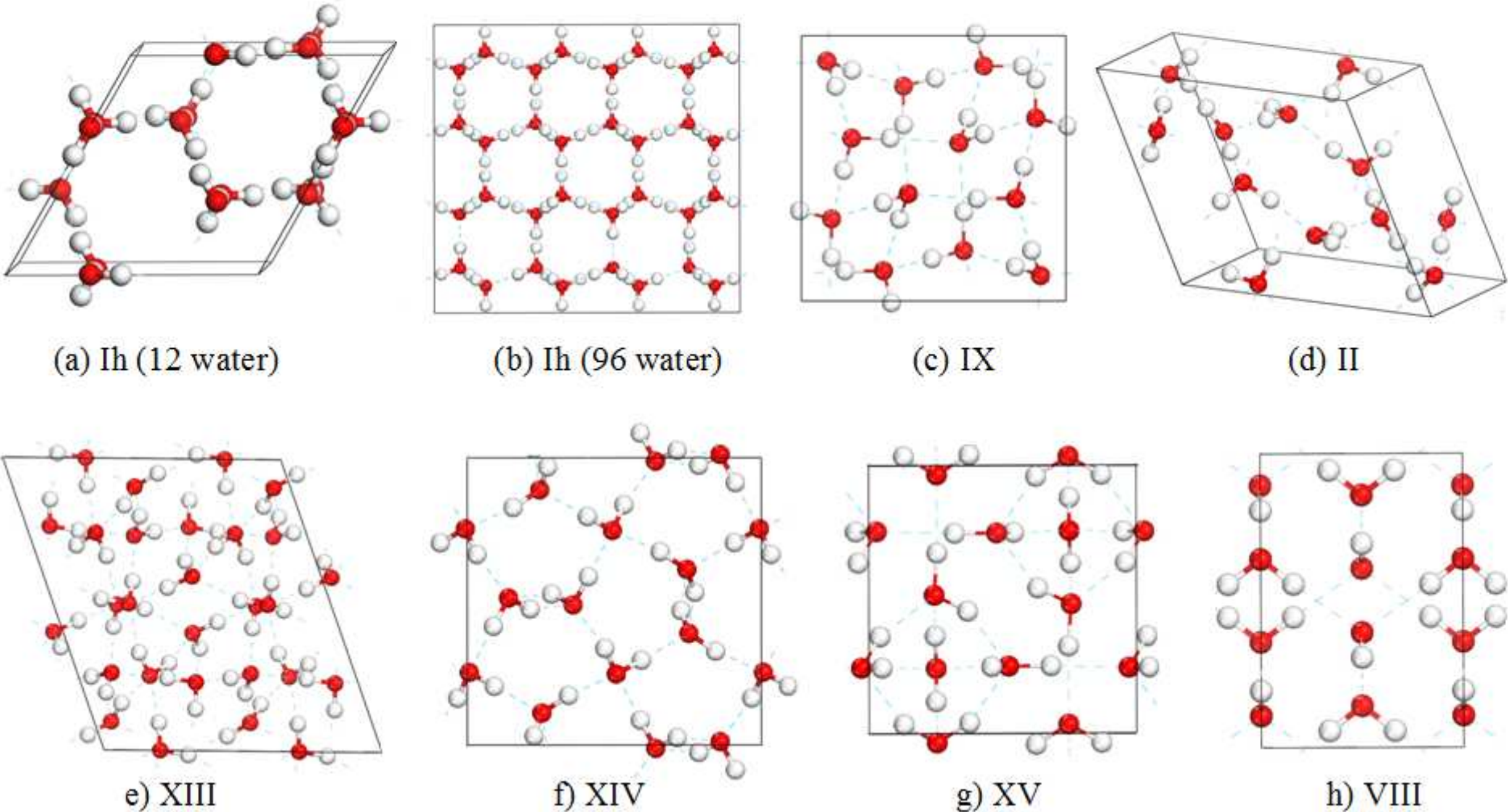}
\caption{\label{fig1} Unit cells of the ice phases (I$h$, IX, II, XIII, XIV, XV, and VIII) studied here. 
The ice I$h$ structure (96 water molecule) is obtained from ref.~\cite{pan_08} and all the proton ordered phases of ice are obtained 
from various scattering experiments.~\cite{ice-XI_96,ice-9_93,ice-2,ice-13-14,ice-15,ice-8}
The optimized coordinates of the ice structures are given in the supplementary material.~\cite{supplement}
}
\end{center}
\end{figure*}

In an earlier study on ice we found that the effects of vdW become increasingly important upon going from the low- to the high-density phases
and capturing this variation in the vdW energy is essential to get the transition pressures between the ice phases within an order 
of magnitude of experiment.~\cite{santra_prl_2011}
Here, we extend on the previous study significantly by
reporting results on the accuracy of the cohesive properties of individual phases of ice obtained from a wide range of vdW inclusive functionals.
Also by looking at the enthalpies of ice as a function of pressure we have obtained a more detailed picture of the 
stability range of each ice phase predicted from the different functionals. 
The approaches used here include:
i) vdW$^{\rm TS}$, which involves an explicit summation of pair-wise (two-body) vdW dispersion interactions among all atom pairs using their
respective vdW $C_6$ coefficients which are functionals of the electron density;~\cite{TS_vdw_09} 
(ii) vdW$^{\rm MB}$, an extension of vdW$^{\rm TS}$ that accounts for electrodynamic screening and 
many-body vdW interactions within the dipole approximation;~\cite{MBD_prl_12} and 
(iii) various functionals from the ``vdW-DF'' family.~\cite{Dion_vdw_04,vdW-DF2,klimes_10}
All vdW-DFs are calculated via a model dynamic response function and long range pairwise approximations.~\cite{Dion_vdw_04}
These various vdW inclusive approaches have been reasonably successful in modeling a wide variety of materials~\cite{langreth_2009,tkatchenko_mrs_2010,klimes_PRB_11,distasio_pnas_12,klimes_jcp_2012,reilly_jpcl_2013}
including different phases of water starting from clusters~\cite{santra_jcp_2008,kelkkanen_jcp_2009,klimes_10} 
to condensed phases.~\cite{santra_prl_2011,murray_prl_2012,wang_jcp_2011,mogelhoj_jpcb_2011,zhang_jctc_2011,zhaofeng_thesis_2012}
In this study we find that
all vdW inclusive functionals considered predict phase transition pressures in much better agreement with experiment
than the functionals which do not include vdW.
%
However, the precise values of the lattice constants and lattice energies are highly sensitive to the choice of
vdW inclusive method. 
Moreover, none of the functionals can simultaneously produce energetics and volumes of the ice phases with high enough precision 
to yield a phase diagram that correctly captures all the phases found in experiments.

In the next section details of the simulation methods are provided. 
This is followed by discussions of our results for the equilibrium lattice energies (\ref{sec3a}), the equilibrium volumes (\ref{sec3b}), 
enthalpies (\ref{sec3c}), and a decomposition of total energies focusing on exchange and correlation energies (\ref{sec3d}). 
Conclusions and a short perspective on future work are given in section~\ref{sec4}.

\section{Simulation Details}
\label{sec2}


We have computed and analyzed the equilibrium lattice energies, volumes, and enthalpies of several ice phases. 
This includes the ambient pressure phase of ice, ice I$h$,
and all the proton ordered high-pressure phases, namely, in order of increasing pressure,
ice IX, II, XIII, XIV, XV, VIII.
We have focused on proton ordered phases because they are more straightforward to model than the proton disordered phases.
%
The initial structures used for the proton ordered phases have been obtained 
from experiment~\cite{ice-XI_96,ice-9_93,ice-2,ice-13-14,ice-15,ice-8}
and the unit cells used are shown in Fig.~\ref{fig1}.
Proton disordered ice I$h$ is modeled with the 12 water unit cell proposed by Hamann.~\cite{hamann_97}
The results obtained from the 12 molecule cell have also been compared to results from a unit cell of 96 water molecules (Ref.~\cite{pan_08,pan_10}).
These results reveal that the lattice energies obtained from the 12 and 96 water molecule unit cells are within 1 meV/H$_2$O and
the equilibrium volumes differ only by $<$0.01 \AA$^3$/H$_2$O with PBE.~\cite{PBE}

%
%
%
The lattice energy per H$_2$O ($\Delta E$) of ice is obtained by subtracting the total energy of $N$ isolated H$_2$O molecules ($E^{\rm H_2O}$) 
from the total energy of the ice unit cell ($E^{\rm Ice}$) containing $N$ molecules of H$_2$O, i.e., 
\begin{equation}
     \Delta E = (E^{\rm Ice}-N\times E^{\rm H_2O})/N  \quad.
\label{eqn_lattice}
\end{equation}
At zero pressure the theoretical equilibrium lattice energies and volumes are obtained
by varying the lattice parameters isotropically within $\pm$20\% of the experimental values 
and fitting the resultant energy-volume curves to the Murnaghan equation of state.~\cite{murnaghan44}
By isotropic variation we mean that the ratios of the lattice parameters are kept fixed at the experimental value,
which is a reasonable approximation that has an insignificant influence on the computed properties.
For example, performing a rigorous test on ice VIII by varying the $c/a$ ratio of the lattice parameters
provides changes of $<$0.5 meV/H$_2$O and $<$0.02 \AA$^3$/H$_2$O, respectively in the equilibrium lattice energy 
and volume when compared to the results obtained by fixing the $c/a$ ratio at the experimental value.~\cite{santra_thesis_2010}
%
%
Also previously it was shown that for ice I$h$ the equilibrium 
$c/a$ ratio is very similar (within $\sim$0.4\%) to the experimental value
when calculated with various \emph{xc} functionals.~\cite{feibelman_08}
%


The properties of the various ice phases have been investigated with seven functionals, representing a number of different classes of functional.
These include, PBE, a widely used GGA functional, and PBE0,~\cite{PBE0} a hybrid exchange variant of PBE.
Neither of these functionals account for vdW forces. 
We have also considered PBE+vdW$^{\rm TS}$ and PBE0+vdW$^{\rm TS}$, vdW inclusive versions of PBE and PBE0,
where the vdW interaction is calculated with the Tkatchenko and Scheffler (TS) scheme.~\cite{TS_vdw_09}
The \ital{xc} energy ($E_{xc}$) in this scheme takes the form 
\begin{equation}
     E_{xc} = E_{x}^{\rm GGA/hybrid} + (E_{c}^{\rm LDA} + E_{c}^{\rm GGA}) + E_{\rm vdW}^{\rm TS} \quad,
\label{eqn_TS}
\end{equation}
where, $E_x^{\rm GGA/hybrid}$ is the PBE or PBE0 exchange, $E_c^{\rm LDA}$ is the LDA correlation, 
$E_c^{\rm GGA}$ is the PBE semi-local correlation correction, and $E_{\rm vdW}^{\rm TS}$ is the vdW energy in the TS scheme.
%
%
In addition, we have employed an extension of the ${\rm vdW}^{\rm TS}$ approach which takes into account
many-body (MB) dispersion and long-range electrostatic screening.~\cite{MBD_prl_12}
In this case $E_{\rm vdW}^{\rm TS}$ in Eq. (1) is replaced with MB dispersion energy terms and the~\emph{xc} functional is 
referred to as PBE0+vdW$^{\rm MB}$.
Another approach to incorporate vdW within DFT is employed here, specifically the approach generally referred to as ``vdW-DF''.~\cite{Dion_vdw_04}
In this case the total~\emph{xc} energy takes the form
\begin{equation}
     E_{xc} = E_{x}^{\rm GGA} + E_{c}^{\rm LDA} + E_{c}^{\rm NL} \quad,
\label{eqn_vdw-df}
\end{equation}
where, $E_x^{\rm GGA}$ is GGA exchange, 
and $E_c^{\rm NL}$ is the nonlocal correlation energy through which the vdW interactions are captured.
We have used three functionals from this category which we refer to as
revPBE-vdW,~\cite{Dion_vdw_04} optPBE-vdW,~\cite{klimes_10} and rPW86-vdW2.~\cite{vdW-DF2}
The difference between revPBE-vdW (originally proposed in ref.~\cite{Dion_vdw_04}) and optPBE-vdW is in the exchange functional only.
The former employs revPBE~\cite{revPBE} exchange, whereas the latter uses optPBE exchange~\cite{klimes_10}
which was developed by fitting interaction energies obtained for the S22 data set.~\cite{s22}
Typically optPBE exchange is less repulsive than revPBE at intermediate and short inter-atomic distances.
Compared to the above two functionals rPW86-vdW2 has a different exchange rPW86~\cite{rPW86_2009} and 
a modified nonlocal correlation functional.~\cite{vdW-DF2}
We should note that all three vdW-DF functionals utilize GGA exchange.

The calculations with PBE and PBE+vdW$^{\rm TS}$ were performed with the all electron 
numeric atom-centered orbital (NAO) basis set code FHI-aims.~\cite{FHI-aims}
Sufficiently large basis sets (``\emph{tier}2'' for H and ``\emph{tier}3'' for O) were employed to calculate total
energies and to optimize structures.
PBE0, revPBE-vdW, optPBE-vdW, and rPW86-vdW2 calculations were done with the VASP code~\cite{vasp-1,vasp-2}
with the hardest projector-augmented wave (PAW) pseudopotentials and a 1000 eV plane-wave basis set cut off.
The $E_{c}^{\rm NL}$ is calculated self-consistently with the efficient algorithm of 
Rom\'{a}n-P\'{e}rez and Soler~\cite{soler_09}
employing 30 interpolation points for the $q_0$ function with a saturation value $q_0^{\rm cut}=10$ a.u.,
as implemented by Klime\v{s}~\emph{et al.} in VASP.~\cite{klimes_PRB_11}
%
These settings are found to be very accurate and details of the implementation and tests performed for a variety of solids
can be found in ref.~\onlinecite{klimes_PRB_11}.
%
%
For all the ice structures the atoms are fully relaxed with all of the~\emph{xc} functionals 
(except with PBE0+vdW$^{\rm TS}$ and PBE0+vdW$^{\rm MB}$)
without any symmetry constraints until all forces are less than 0.01 eV/\AA.
The energy-volume curves of each ice phase with PBE0+vdW$^{\rm TS}$ and PBE0+vdW$^{\rm MB}$ 
were produced by performing single point energy calculations on the PBE0 optimized geometries at 
different volumes.
%
%
For the calculations of any GGA exchange based functional the number of~\bd{k} points are chosen 
so that the spacing in the~\bd{k} point grid
in each direction of reciprocal space is within 0.02~\AA$^{-1}$ to 0.04~\AA$^{-1}$
for all of the ice phases.
For the hybrid functional (PBE0) calculations the number of~\textbf{k} points
are doubled in each direction compared to the GGA calculations,
which provides total energies converged to within $<$1 meV/H$_2$O.
With VASP the energy of the water monomer was calculated within a cubic cell of length 20~\AA. 
%

\section{Results}

\begin{figure}
\begin{center}
    \includegraphics[width=8cm]{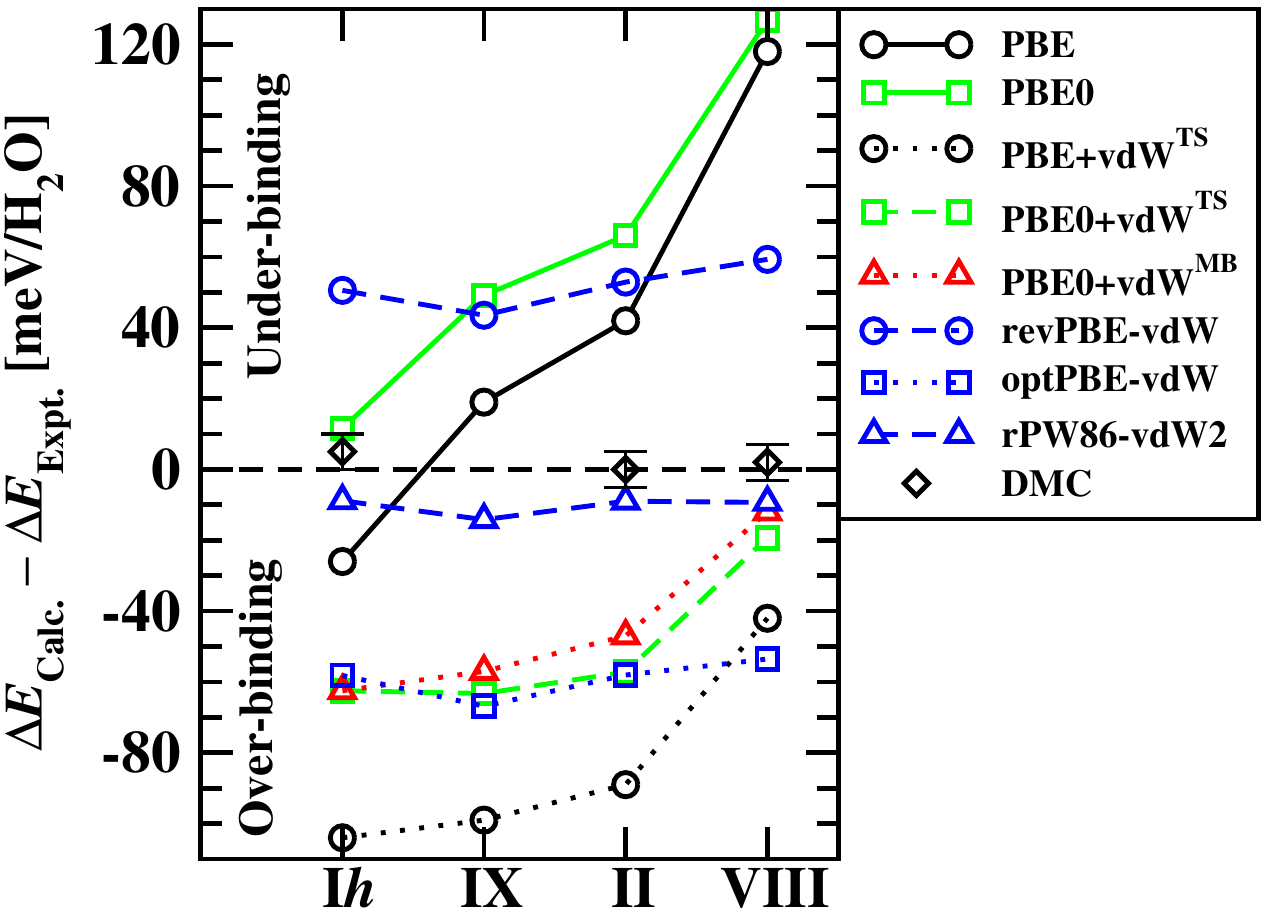}
    \caption{\label{fig2} Differences in the lattice energies of ice calculated ($\Delta E_{{\rm Calc.}}$) with various DFT functionals 
and DMC~\cite{santra_prl_2011} compared to experiment ($\Delta E_{{\rm Expt.}}$). Zero on the Y axis indicates perfect agreement with experiment.}
\end{center}
\end{figure}

\begin{table*}
\caption{\label{tb1} Equilibrium lattice energies of different ice phases with various methods.
The relative lattice energies of the high-pressure ice phases with respect to ice I$h$ are given in parenthesis.
All energies are in meV/H$_2$O.}
\begin{ruledtabular}
\begin{tabular}{c|ccccccc}
\hline
           & I$h$      & IX        & II        & XIII      & XIV       & XV         & VIII     \\
\hline
Expt.$^a$
           & -610 (0)  & -606 (5)  & -609 (1)  &  --       &   --      &   --       & -577 (33) \\
DMC$^b$
           & -605 (0)  & --        & -609 (-4) &  --       &   --      &   --       & -575 (30) \\
\hline
PBE        & -636 (0)  & -587 (49) & -567 (69) & -556 (80) & -543 (93) & -526 (110) & -459 (177)\\
PBE0       & -598 (0)  & -557 (41) & -543 (55) & -530 (67) & -518 (80) & -504 (94)  & -450 (148)\\
\hline
PBE+vdW$^{\rm TS}$
           & -714 (0)  & -705 (9)  & -698 (16) & -695 (19) & -690 (24) & -678 (36)  & -619 (95) \\
PBE0+vdW$^{\rm TS}$
           & -672 (0)  & -670 (2)  & -666 (6)  & -661 (11) & -656 (16) & -646 (26) & -596 (76)  \\
PBE0+vdW$^{\rm MB}$
           & -672 (0)  & -663 (9)  & -656 (16) & -648 (22) & -642 (30) & -629 (43) & -589 (83)  \\
\hline
revPBE-vdW & -559 (0)  & -563 (-4) & -556 (3)  & -555 (4)  & -552 (7)  & -545 (14) & -517 (42)  \\
optPBE-vdW & -668 (0)  & -673 (-5) & -667 (1)  & -666 (2)  & -664 (4)  & -656 (12) & -630 (38)  \\
rPW86-vdW2 & -619 (0)  & -621 (-2) & -618 (1)  & -615 (4)  & -605 (14) & -605 (14) & -586 (33)  \\
\hline\hline
\multicolumn{6}{l}{
$^a$Ref.~\onlinecite{whalley_jcp_84};
$^b$The DMC statistical error bar is $\pm$5 meV/H$_2$O (Ref.~\onlinecite{santra_prl_2011})
}\\
\end{tabular}
\end{ruledtabular}
\end{table*}

In this section we report how the above mentioned DFT~\ital{xc} functionals describe the different phases of ice
by examining equilibrium lattice energies, volumes at zero pressure and the relative enthalpies of the various phases.
Subsequently we report an analysis of the individual contributions from exchange and correlation to the lattice energies.

\subsection{Lattice energies at zero pressure}
\label{sec3a}

The lattice energy is one of the key characteristic quantities of a solid and we use it here 
to evaluate the performance and deficiencies of different~\emph{xc} functionals in describing ice.
Previously most analysis of lattice energies concentrated on ice I$h$,~\cite{hamann_97,feibelman_08,hamada_10}
however, recently we showed that studying ice I$h$ alone is not enough to establish the general behavior of an~\emph{xc} functional
over the entire phase diagram of water.~\cite{santra_prl_2011}
%
Here we have calculated the lattice energies of different ice phases using a wide variety of~\emph{xc} functionals
and made comparisons with experiments~\cite{whalley_jcp_84,rottger_Ih_94,ice-9_93,ice-13-14,ice-15}
and diffusion quantum Monte Carlo (DMC)~\cite{santra_prl_2011} whenever possible (Table~\ref{tb1}).
We note that the DFT and DMC lattice energies reported in Table~\ref{tb1} do not include nuclear zero-point energies (ZPEs)
and are directly comparable with the experimental lattice energies provided by Whalley,~\cite{whalley_jcp_84}
in which ZPE contributions were removed and the energies were extrapolated to 0 K. 
%
%
In Fig.~\ref{fig2} the differences in the calculated and experimental lattice energies are shown for ice I$h$, IX, II, and VIII
for all the functionals considered.
It can be seen that for the phases for which DMC data is available
the agreement between DMC and experiment is excellent, differing only by 5 meV/H$_2$O at most,
which is also the size of the DMC statistical errors.

\begin{figure*}
\begin{center}
    \includegraphics[width=14cm]{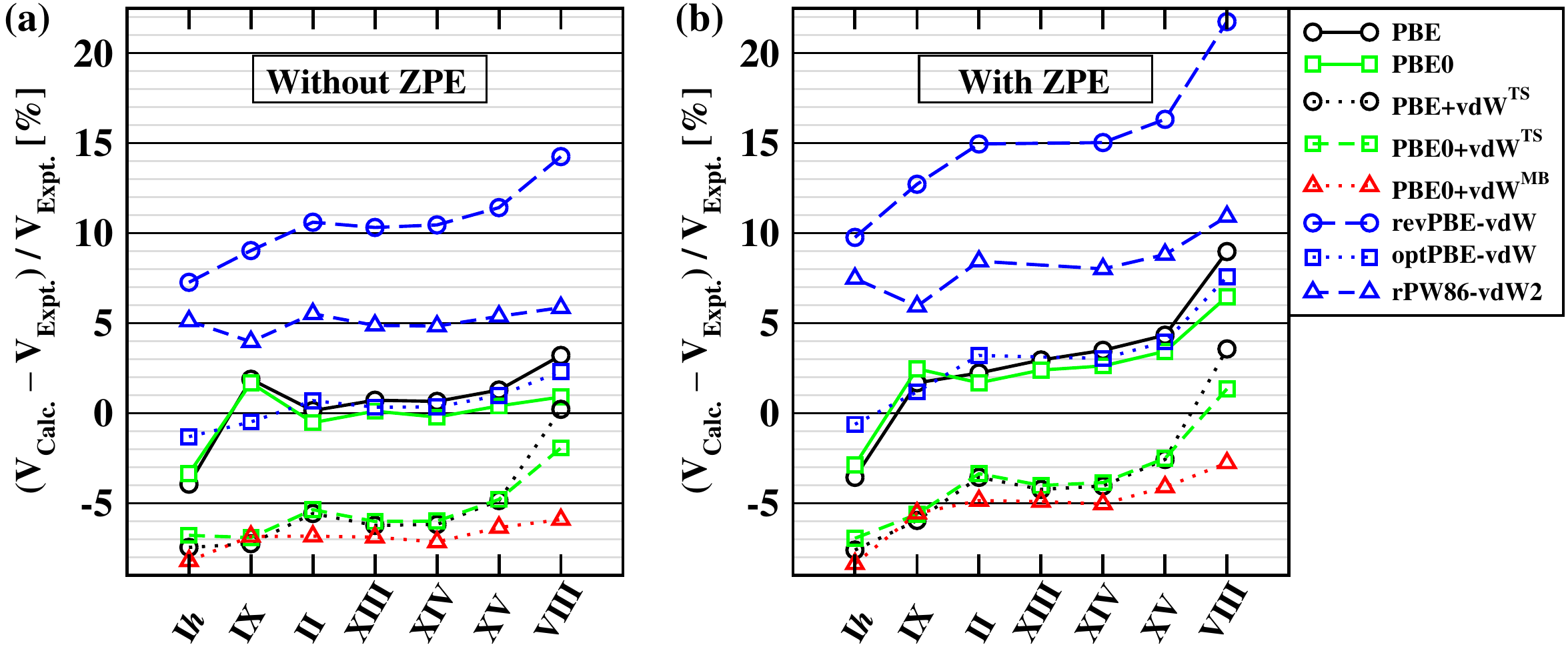}
\caption{\label{fig3} Percentage differences in the calculated
equilibrium volumes (V$_{\rm Calc.}$) compared to experiment (V$_{\rm Expt.}$) (a) without and (b) with zero point energies (ZPE) in V$_{\rm Calc.}$.
The zero value on the Y axis designates the experimental reference values. Here positive errors indicate 
larger computed volumes and negative errors indicate smaller computed volumes compared to experiment.}
\end{center} 
\end{figure*}

We begin with the performance of functionals which do not account for vdW (PBE and PBE0) on the absolute values of the lattice energies.
%
The behavior of PBE for ice I$h$ is well known, it over-estimates the lattice energy: overestimations of between 30 to 100 meV/H$_2$O
have been reported depending on the computational set-up used (mainly the quality of the basis sets and pseudopotentials).~\cite{pan_08,hamada_10,carrasco_prl_2011,murray_prl_2012}
Here, using a full potential all-electron approach and very tightly converged NAO basis sets, 
we find that PBE over-estimates the lattice energy of ice I$h$ by $\sim$26 meV/H$_2$O (Fig.~\ref{fig2}). 
This is in close agreement with the results from highly converged PAW calculations reported in ref.~\onlinecite{feibelman_08}. 
Interestingly the established notion that PBE overestimates the lattice energies of ice does not hold
for the high-density phases. 
For example, PBE exhibits a $\sim$125 meV/H$_2$O underestimation for the lattice energy of ice VIII.
The story is somewhat similar for PBE0, the hybrid variant of PBE.
PBE0 predicts a very good lattice energy for ice I$h$ (only 15 meV/H$_2$O less than experiment) but simultaneously underestimates the lattice energy 
of the high-density ice phases.
%
%
We believe that the behavior seen here for PBE and PBE0 is likely to apply to many other GGA and hybrid~\emph{xc} functionals.
For example, our calculations show that BLYP and revPBE GGA functionals under-estimate the lattice energy of ice VIII by 246 meV/H$_2$O and 316 meV/H$_2$O, respectively.
Similar findings have also been reported for B3LYP.~\cite{pccp_ice_12}
%
%
%

In general we find that 
when vdW is accounted for the differences between the calculated and experimental lattice energies are much less sensitive
to the particular phases being examined (Fig.~\ref{fig2}).
%
Considering first the vdW$^{\rm TS}$ scheme, with the
PBE+vdW$^{\rm TS}$ and PBE0+vdW$^{\rm TS}$ functionals the lattice energies are on average
$\sim$100 meV/H$_2$O and $\sim$60 meV/H$_2$O, respectively too large compared to experiment.
The smaller errors obtained from PBE0+vdW$^{\rm TS}$ largely arise from the difference
between PBE and PBE0 since the vdW contributions from these functionals obtained with the TS scheme 
are similar (to within 15 meV/H$_2$O).
The contributions from vdW interactions beyond two-body vdW$^{\rm TS}$ to lattice energies are found to be small for the phases considered.
%
%
%
Specifically, the PBE0+vdW$^{\rm MB}$ approach reduces the error by 7-17 meV/H$_2$O for the high-density phases compared to the standard PBE0+vdW$^{\rm TS}$.
%
%
%
%
%
A noticeable exception to the consistent performance of vdW$^{\rm TS}$ found for ice I$h$, IX, and II is the
highest density ice VIII phase (by 40-50 meV/H$_2$O [Fig.~\ref{fig2}]).
%
%
This inconsistency is largely due to the shortcomings of the damping function used in the vdW$^{\rm TS}$ approach 
in describing the interpenetrating H bond network in ice VIII which has water molecules that do not form H bonds with each other
as close as 2.9~\AA\ apart. 
%

When the vdW-DF functionals are utilized the errors are more consistent for the phases (Fig.~\ref{fig2}). 
However, the magnitude and sign of the error varies considerably from one functional to another, e.g.,
on average optPBE-vdW produces too large ($\sim$60 meV/H$_2$O) and revPBE-vdW produces too small ($\sim$50 meV/H$_2$O) lattice energies
compared to experiment.
The fact that revPBE-vdW underestimates the lattice energy is not a surprise and consistent with results obtained with this functional
for small molecules and water clusters.~\cite{gulans_prb_2009,kelkkanen_jcp_2009,klimes_10,vdW-DF2,klimes_jcp_2012}
Previously reported lattice energies of ice I$h$ with revPBE-vdW are 30-35 meV/H$_2$O larger than what we obtain here. 
This is not a very substantial difference and we suspect
it is mostly down to differences in pseudopotentials.~\cite{hamada_10,murray_prl_2012}
Here, rPW86-vdW2 provides the best agreement with experimental lattice energies 
being consistently within $\sim$15 meV/H$_2$O of experiment for all ice phases.
%
%

\begin{table*}
\caption{\label{tb2} Comparisons of the calculated and experimental equilibrium volumes (\AA$^3$/H$_2$O) of the various ice phases.
MAE is mean absolute error (\%) and ME is mean error (\%) (averaged over all the ice phases) with respect to the experimental volumes.
Errors with and without zero point vibration (ZPE) are shown.  
For the MAEs the positive sign indicates larger volumes and the negative sign smaller volumes compared to experiment.} 
\begin{ruledtabular}
\begin{tabular}{c|ccccccc|cc|cc}
\hline
 & & & & & & & & \multicolumn{2}{c|}{Without ZPE} & \multicolumn{2}{c}{With ZPE} \\
           & I$h$        & IX        & II        & XIII      & XIV       & XV        & VIII      & MAE  & ME    & MAE & ME \\
\hline
Expt.      & 32.05$^a$ & 25.63$^b$ & 24.97$^c$ & 23.91$^d$ & 23.12$^d$ & 22.53$^e$ & 20.09$^c$ & --   & --    & --  & -- \\
DMC$^f$        
           & 31.69     &           & 24.70     &           &           &           & 19.46     & --   & --    & --  & -- \\      
\hline
PBE        & 30.79     & 26.11     & 25.01     & 24.08     & 23.27     & 22.82     & 20.74     & 1.69 & 0.57  & 4.00 & 2.99 \\
PBE0       & 30.98     & 26.06     & 24.84     & 23.94     & 23.07     & 22.62     & 20.27     & 1.03 & -0.14 & 3.14$^\star$ &2.32$^\star$\\
\hline
PBE+vdW$^{\rm TS}$
           & 29.67     & 23.86     & 23.62     & 22.44     & 21.71     & 21.47     & 20.13     & 5.52 & -5.47 & 4.51 & -3.49  \\
PBE0+vdW$^{\rm TS}$
           & 29.88     & 23.85     & 23.63     & 22.47     & 21.74     & 21.45     & 19.70     & 5.39 & -5.39 & 4.05$^\star$ & -3.05$^\star$\\
PBE0+vdW$^{\rm MB}$
           &  29.42    & 23.87     & 23.26     & 22.26     & 21.45     & 21.10     & 18.90     & 6.88 & -6.88 &  5.08$^\star$ & -5.08$^\star$ \\
\hline
revPBE-vdW &  34.38    & 27.94     & 27.62     & 26.38     & 25.54     & 25.10     & 22.96     & 10.27 & 10.27& 15.09 & 15.09 \\
optPBE-vdW &  31.63    & 25.50     & 25.15     & 23.99     & 23.20     & 22.75     & 20.55     & 0.92  & 0.42 & 3.21 & 3.09 \\
rPW86-vdW2 &  33.69    & 26.65     & 26.35     & 25.07     & 24.24     & 23.74     & 21.27     & 5.09  & 5.09 & 8.22 & 8.22 \\
\hline \hline
\multicolumn{10}{l}{
$^a$10 K, Ref.~\onlinecite{rottger_Ih_94}; 
$^b$30 K Ref.~\onlinecite{ice-9_93};
$^c$0 K Ref.~\onlinecite{whalley_jcp_84}; 
$^d$80 K Ref.~\onlinecite{ice-13-14}; 
$^e$80 K Ref.~\onlinecite{ice-15};}\\
\multicolumn{12}{l}{$^f$The DMC statistical errors are $\pm$0.01, $\pm$0.20, $\pm$0.02 \AA$^3$/H$_2$O, respectively for ice I$h$, II, VIII} \\
\multicolumn{10}{l}{(Ref.~\onlinecite{santra_prl_2011}); $^\star$See Ref.~\onlinecite{note_on_PBE0+vdW}} \\
\end{tabular}
\end{ruledtabular}
\end{table*}

%
The role played by vdW interactions in the phase diagram of ice is most evident when relative energies between the ice phases are considered.
Both the vdW$^{\rm TS}$ and vdW-DF approaches provide results which are in much closer agreement with experiment and DMC 
than the standard GGA and hybrid functionals in this regard.
Table~\ref{tb1} shows that experimentally the energy difference between ice I$h$ and the highest density ice VIII phase is 33 meV/H$_2$O.
Although the relative stabilities of ice XIII, XIV, and XV are not known (either from experiment or DMC) they should also fall within the 33 meV/H$_2$O window
since ice VIII is the least stable phase at zero pressure and 0 K (of all the phases studied here). 
However, when calculated with PBE and PBE0 the energy difference between ice I$h$ and ice VIII is far too large ($>$140 meV/H$_2$O).
Likewise, when comparison with experiment is possible the phases between ice I$h$ and ice VIII are also destabilized too much. 
%
All vdW inclusive functionals reduce the energy differences between the phases, bringing them into much closer agreement with experiment.
For example, the energy difference between ice I$h$ and ice VIII comes down to only 76 meV/H$_2$O and 33 meV/H$_2$O with PBE0+vdW$^{\rm TS}$ and rPW86-vdW2, respectively.

As noted earlier, ZPE contributions were not considered in the above discussions as ZPE effects do not play a significant 
role in determining the relative energies of the various ice phases.
The main effect from ZPE contributions is to reduce the lattice energies by about 120-110 meV/H$_2$O.
This applies across the board for all phases and functionals considered, although a small monotonic decrease on the level of 6-10 meV/H$_2$O 
is seen upon going from the low- to the high-density phases.
However, as we will see in the next section ZPE effects do influence the equilibrium volumes significantly.

\subsection{Equilibrium volumes}
\label{sec3b}

%
The equilibrium volume, which is a measure of the density of the phases, is
another important quantity used to assess the performance of different theoretical methods.
Previous efforts focused mostly on calculating the density of ice I$h$,  
whereas here we seek to understand how functionals perform for a range of phases.
%
Table II and Fig.~\ref{fig3} show comparisons of the calculated and experimental volumes.
The equilibrium volumes of ice I$h$ obtained using both PBE and PBE0 are $\sim$4\% smaller than experiment, 
which is consistent with previous calculations.~\cite{feibelman_08,murray_prl_2012}
%
%
Upon going to higher densities, however, contrasting behavior is observed and as one moves to higher densities there is an increasing
tendency to overestimate the volume.
Indeed for ice VIII the volume is overestimated by 4\% with both PBE and PBE0. 
%
Clearly, the greater underestimation of vdW interactions at higher densities leads to a progressive overestimation of the equilibrium volumes. 
%
%
Overall though, and in contrast to the lattice energies,
the performance of PBE and PBE0 for the equilibrium volumes are reasonable, differing by $<$2\% from experiment when averaged over all ice phases.

\begin{figure*}
\begin{center}
    \includegraphics[width=14cm]{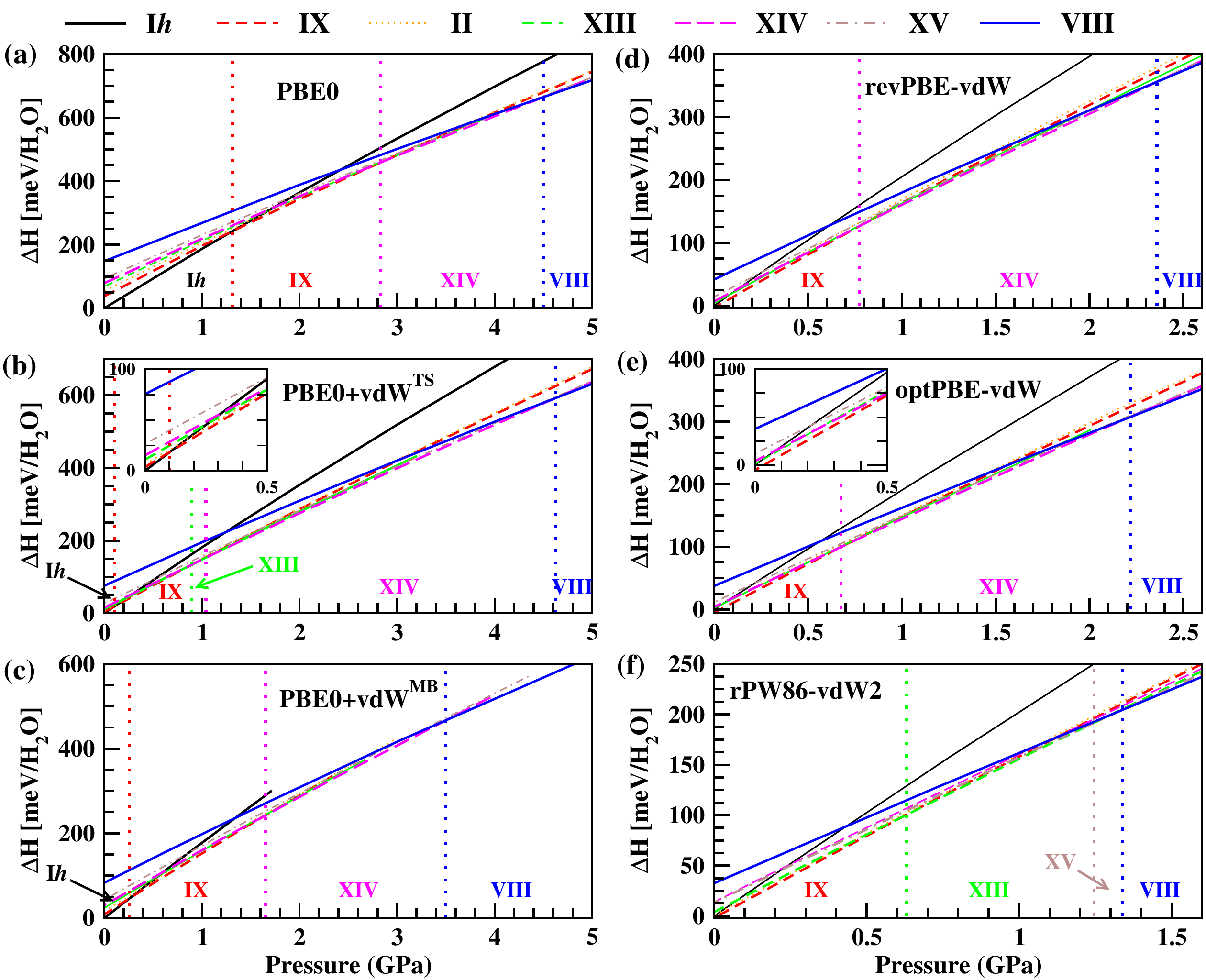}
\caption{\label{fig4} Change in the enthalpies ($\Delta$H) of the ice phases 
relative to the enthalpy of ice I$h$ at T=0 and P=0 calculated with 
(a) PBE0, (b) PBE0+vdW$^{\rm TS}$, (c) PBE0+vdW$^{\rm MB}$, (d) revPBE-vdW, (e) optPBE-vdW, and (f) rPW86-vdW2.
The vertical dotted lines indicate the transition pressures. 
The most stable ice phases along the pressure axis are indicated at the bottom of each panel.
The insets show elaborations of the PBE0+vdW$^{\rm TS}$ and optPBE-vdW plots within the 0.0-0.5 GPA pressure range.}
\end{center}
\end{figure*}

With vdW$^{\rm TS}$ the volumes are decreased by 3-9\% from 
their parent functionals (PBE and PBE0) and in comparison to experiment the volumes actually become worse.
The underestimated volumes obviously correlate with the underestimated 
lattice energies predicted by these approaches (Fig.~\ref{fig2}).
Going beyond two-body vdW$^{\rm TS}$ the equilibrium volumes are reduced further and compared to experiment the average difference becomes $\sim$7\%.  
It is noteworthy that 
vdW$^{\rm MB}$ reduces the volume of ice VIII more than vdW$^{\rm TS}$, and as a result it improves the relative change in the volume
with respect to ice XV, which in turn substantially affects the phase transition pressures (section~\ref{sec3c}).
Contrary to the performance of the vdW$^{\rm TS}$ approaches, the equilibrium volumes obtained from revPBE-vdW are on average $\sim$10\% too large (Fig.~\ref{fig3}).
%
Such behavior of revPBE-vdW has been attributed to the overly repulsive revPBE exchange functional and is analogous to what 
has been found earlier with this functional for many other solids.~\cite{klimes_PRB_11,klimes_jcp_2012}
The rPW86-vdW2 functional incorporates improvements in both the exchange and correlation components of the functional compared to revPBE-vdW 
and we find that this is reflected in improved volumes, being on average $\sim$5\% larger than experiment. 
%
%
However, this performance is still inferior to PBE and PBE0.
Of the vdW inclusive functionals optPBE-vdW provides the smallest average error being within 2\% of experiment for all the ice phases.

Unlike the lattice energies, the effects of ZPE on the equilibrium volumes cannot be ignored, especially in the higher density phases.
%
%
The effect of ZPE on the equilibrium volumes is estimated by computing free energy as a function of volume as 
$F(V)=E(V)+\frac{1}{2}\sum_\nu \hbar \omega_\nu(V)$,
where $\omega_\nu(V)$ being the frequency of phonon mode $\nu$ at a given volume. 
In line with the recent study of Murray and Galli,~\cite{murray_prl_2012} we find that with PBE the volumes of ice I$h$ and VIII
increase by $\sim$0.5\% and $\sim$5.5\%, respectively when ZPE effects are accounted for (Fig.~\ref{fig3}(b) and Table~\ref{tb2}).
%
%
Indeed overall we find that the ZPE effects gradually increase from the low- to high-density phases
and depending on the functional the increase in the equilibrium volume 
for the highest density ice VIII phase is somewhere between 3-6\%.~\cite{note_on_PBE0+vdW}
%
%
Thus, compared to experiments the mean absolute error in predicting volumes of phases increases by $\sim$2\% for vdW-DF 
and decreases by $\sim$1.5\% for vdW$^{\rm TS}$ when ZPE effects are accounted for.
%
%
Overall for all the vdW inclusive functionals, when ZPE effects are taken into account
optPBE-vdW and PBE0+vdW$^{\rm TS}$ are the two best functionals in terms of predicting volumes (Table~\ref{tb2}).

\subsection{Enthalpy}
\label{sec3c}

%
Apart from absolute lattice energies and densities,
accurate predictions of phase transitions are important if a functional is to be of real value in exploring the phase diagram of water.
%
%
In ref.~\cite{santra_prl_2011} we showed that vdW interactions had a huge impact on the predicted phase transition pressures 
between the various phases of ice considered.
Here we extend this study by calculating the enthalpies of different phases to establish the most stable phases at different pressures 
predicted by the various~\emph{xc} functionals.
%
%
Pressures, P(V), at different volumes have been calculated from the Murnaghan equation of state:~\cite{murnaghan44}
\begin{equation}
 P(V) = B_0/B_0^\prime ((V_0/V)^{B_0^\prime}-1),
\end{equation}
where $B_0$, $B_0^\prime$, $V_0$ are the equilibrium bulk modulus, the derivative of 
the bulk modulus with respect to pressure, and the equilibrium volume at zero pressure, respectively.
Fig.~\ref{fig4} shows the enthalpies of the various phases as a function of pressure relative to the enthalpy of ice I$h$ (at 0 K and zero pressure).
%
%
The most stable phase at each pressure is the one with the lowest enthalpy and the crossovers between different phases indicate
the pressures at which the phase transitions are predicted to occur.
The phase transition pressures predicted from all functionals are also summarized in Table~\ref{tb3}.
According to the most recent experimental phase diagram of water, upon pressurizing the ambient pressure ice I$h$ phase the high-pressure  
proton ordered phases are expected to occur in the following sequence: ice IX, II, XIII, XIV, XV, and VIII.~\cite{salzmann_pccp_2011}
%
%
However, the exact phase boundaries between these phases have not been determined directly from experiment,
especially when considering the low temperature regime.~\cite{ice_book,ice-15,salzmann_rsc_2007,salzmann_pccp_2011}
%
%
Specifically, between ice I$h$ and IX there is no measured phase boundary available and a reasonable choice is to consider 
the known phase coexistence line between ice I$h$ 
and III, the proton disordered counterpart of ice IX, which appears at $\sim$0.1-0.2 GPa.~\cite{kell_jcp_1968,salzmann_pccp_2011}
The experimental phase boundaries between ice IX, II, and XIII are also unknown, however, they certainly should appear in the pressure window of 0.2-0.8 GPa
since at higher pressures ($\sim$1.2-1.4 GPa) ice XIV and XV are found to be stabilized.~\cite{ice-13-14,ice-15,salzmann_pccp_2011}
The highest density phase, ice VIII, can be found at 1.5 GPa.~\cite{salzmann_pccp_2011}
Now we will discuss how our calculated phase transition pressures compare with the experimental data. 

\begin{table}
\caption{\label{tb3} 
Comparisons of the calculated and experimental transition pressures. 
Only positive transition pressures are reported. All pressures are in GPa.}
\begin{ruledtabular}
\begin{tabular}{c|ccccc}
\hline
	   & IX           & II  & XIII & XIV/XV & VIII \\
\hline
Expt.      & 0.1--0.2$^a$ & --  & 0.2--0.8$^b$ & 1.2--1.4$^b$ & 1.50$^b$ \\
		      
\hline                
PBE        & 1.66         & --  & --   & 3.45  & 6.08 \\
PBE0       & 1.32         & --  & --   & 2.83  & 4.50 \\ 
\hline                     
PBE+vdW$^{\rm TS}$         
	   & 0.26         & --  & --   & 1.25  & 6.37 \\
PBE0+vdW$^{\rm TS}$        
	   & 0.10         & --  & 0.89 & 1.04  & 4.62 \\
PBE0+vdW$^{\rm MB}$        
	   & 0.26         & --  & --   & 1.65  & 3.50 \\
\hline                     
revPBE-vdW & --           & --  & --   & 0.78  & 2.36 \\
optPBE-vdW & --           & --  & --   & 0.68  & 2.22 \\
rPW86-vdW2 & --           & --  & 0.63 & 1.25  & 1.34 \\
\hline \hline
\multicolumn{6}{l}
{$^a$
Ref.~\onlinecite{kell_jcp_1968}; 
$^b$Ref.~\onlinecite{salzmann_pccp_2011}
}\\
\end{tabular}
\end{ruledtabular}
\end{table}

%
%
Table~\ref{tb3} shows that the phase transition pressures obtained from PBE are much too high compared to experiment;
about an order of magnitude too high for ice IX and 3-4 times too high for ice XIV and VIII. 
Small improvements arise using PBE0 with 20-30\% reductions in the transition pressures. 
%
%
The predicted order in which the ice phases appear (I$h$, IX, XIV, and VIII) with increasing pressure agrees with experiment.
However, ice II, XIII, and XV are missing from the PBE (and PBE0) phase diagram,  
i.e., at no positive external pressure do these phases have the lowest enthalpy (Fig.~\ref{fig4}).

%
%
With the vdW inclusive functionals the transition pressures are lowered substantially and are in reasonable agreement with experiment (Table~\ref{tb3}). 
For ice IX, XIII, and XIV the transition pressures obtained from PBE0+vdW$^{\rm TS}$ are within the range of experimental values.
However, PBE0+vdW$^{\rm TS}$ fails to reduce the transition pressure of ice VIII
mainly because the relative lattice energy of ice VIII with respect to ice XV ($\sim$50 meV/H$_2$O) does not improve with vdW$^{\rm TS}$ (Table~\ref{tb1}).
Inclusion of many-body vdW decreases the energy difference between ice VIII and ice XV by 10 meV/H$_2$O 
and brings the relative change in the equilibrium volume (2.2 \AA$^3$/H$_2$O) into better agreement with experiment (2.4 \AA$^3$/H$_2$O). 
Both improvements help in reducing the calculated transition pressure of ice VIII to 3.5 GPa which is closer to the experimental pressure of 1.5 GPa.
Despite the improvements in the transition pressures both vdW$^{\rm TS}$ and vdW$^{\rm MB}$ fail to predict the presence of all of the 
experimentally characterized ice phases on the phase diagram (Fig.~\ref{fig4}).

%
%
The phase diagrams obtained with the three vdW-DFs are not particularly impressive either.
None of the functionals find the ice I$h$ to IX transition at a positive pressure, because they predict ice IX is to be energetically 
more stable than ice I$h$ at zero pressure.
However, the predicted transition pressures for the higher density phases (ice XIII and beyond) are in good agreement
with experiment, differing by no more than a factor of 2.  
Interestingly, since ice XIV and XV are isoenergetic with rPW86-vdW2 (Table~\ref{tb1}) this functional
predicts ice XV to be more stable than ice XIV at all pressures (Fig.~\ref{fig4}(f)). 
For ice VIII all three vdW-DFs reproduce the experimental transition pressure (1.5 GPa) with reasonable accuracy, 
rPW86-vdW2 being the closest (1.34 GPa) followed by optPBE-vdW (2.22 GPa) and then revPBE-vdW (2.36 GPa). 
%


\subsection{Decomposition of the Exchange and Correlation Contributions to the Lattice Energy}
\label{sec3d}

\begin{figure*}
\begin{center}
    \includegraphics[width=14cm]{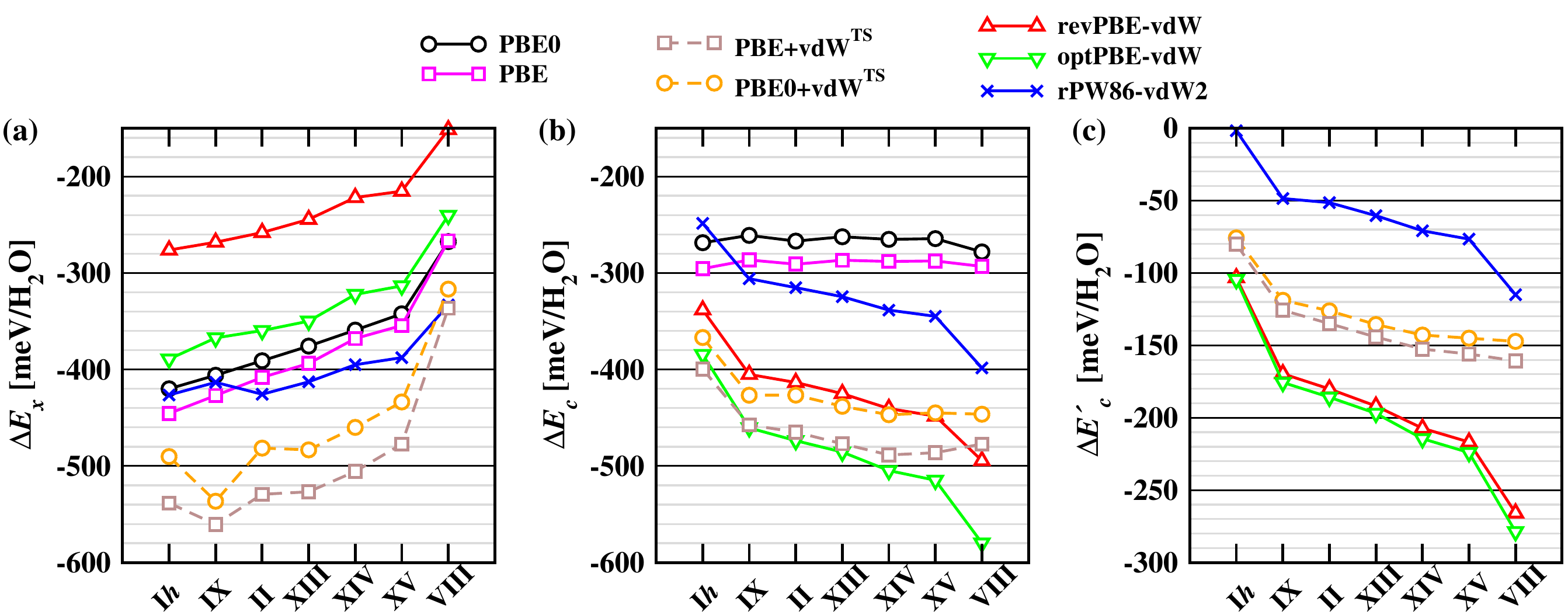}
    \caption{\label{fig5} Contributions to the lattice energies of the various ice phases from (a) the exchange energy ($\Delta{E}_{x}$) [\emph{c.f.} Eq.~\ref{eqn_decomp}], 
(b) the total correlation energy ($\Delta{E}_{c}$), and 
(c) the correlation energy beyond GGA-PBE correlation ($\Delta{E}^{\prime}_{c}$).
All energies have been calculated on the 
equilibrium densities obtained from each functional.
}
\end{center}
\end{figure*}

\begin{table}
\caption{\label{tb4} 
The calculated molecular $C_6$ coefficients of water molecules in the various ice phases and an isolated H$_2$O molecule.
The coefficients are given in Hartree$\cdot$Bohr$^6$. 
}
\begin{ruledtabular}
\begin{tabular}{c|cccc}
\hline
	&revPBE-	&optPBE-	&rPW86-    	&PBE+             \\	
	&vdW	        &vdW	        &vdW2    	&vdW$^{\rm TS}$    \\	
\hline
 Isolated H$_2$O$^a$
        &58.42	        &58.33	        &20.85	        &43.96  \\	
\hline
 I$h$	&61.53		&59.30		&28.55		&48.40	\\
 IX	&64.31		&61.41		&33.33		&51.36	\\
 II	&63.88		&60.49		&33.35		&52.24	\\
 XIV	&64.25		&60.60		&34.67		&53.32	\\
 XV	&64.67		&60.72		&35.28		&53.86	\\
 VIII	&64.40		&59.19		&36.53		&55.86	\\
\hline \hline
\multicolumn{5}{l}
{$^a$The corresponding experimental value is 45.29~\onlinecite{margoliash_jcp_1978}. } \\
\end{tabular}
\end{ruledtabular}
\end{table}


GGA, hybrid-GGA, and vdW inclusive functionals lead to varied results for the ice phases considered.
%
In order to shed more light on why this is
we have decomposed the contributions from exchange and correlation energies 
to the lattice energies for all the~\emph{xc} functionals studied. 
The contribution from the exchange energy ($\Delta E_x$) to the lattice energy is 
obtained by subtracting the exchange energy of $N$ isolated H$_2$O molecules ($E^{\rm H_2O}_x$) 
from the exchange energy of the ice unit cells ($E^{\rm Ice}_x$) containing $N$ H$_2$O molecules 
and can be defined as: 
\begin{equation}
     \Delta E_x = (E^{\rm Ice}_x - N\times E^{\rm H_2O}_x)/N \quad.
\label{eqn_decomp}
\end{equation}
An equivalent definition is used to extract the contribution form the correlation energy ($\Delta E_c$) to the lattice energy.

Figs.~\ref{fig5}(a) and~\ref{fig5}(b) show the variations in $\Delta E_x$ and $\Delta E_c$ for all ice phases  
at the equilibrium densities obtained from each~\emph{xc} functional.
In general we find that upon going from the low to the high density phases the 
 energetically favorable exchange contribution to the lattice energy decreases, just as the lattice energies do.
 %
%
%
%
%
%
%
We also find that the exchange contribution to the lattice energy strongly depends on the equilibrium volumes obtained with the various functionals.
Consequently revPBE-vdW predicts the largest volumes and smallest $\Delta E_x$ and PBE+vdW$^{\rm TS}$ the smallest volumes and largest $\Delta E_x$ for all phases.
%
%
%
Since the hybrid PBE0 exchange yields accurate electrostatic properties (e.g., polarizability, dipole moment, electronic band gap)
for the gas and condensed phases of water~\cite{hammond_09,santra_jcp_2009,jordan_10,labat_11}
it's somewhat useful to consider  PBE0 $\Delta E_x$ as a reference against which we compare other exchange functionals.  
In this regard PBE and optPBE follow PBE0 fairly closely over the entire pressure range.
On the other hand, while rPW86-vdW2 $\Delta E_x$  is within 10 meV/H$_2$O of the PBE0 value for ice I$h$ and IX, it deviates substantially ($>$70 meV/H$_2$O more negative)
for the higher density ice phases, implying that rPW86 exchange is over stabilized compared to PBE0 exchange for the highest pressure phase.
%
%
%
%
Similarly PBE+vdW$^{\rm TS}$ and PBE0+vdW$^{\rm TS}$ $\Delta E_x$ are substantially more negative than PBE0 exchange for all phases except ice VIII. However, this is mainly due to the smaller volumes predicted by these approaches compared to PBE0.
%

%
%

The contributions from correlations to the lattice energies, $\Delta E_c$, show why PBE and PBE0 perform so poorly for the high-density phases.
Specifically we find that the PBE $\Delta E_c$ is nearly constant for all the ice phases, which is
in stark contrast to the predictions from the vdW inclusive~\emph{xc} functionals
that $\Delta E_c$ increases from the low to high density phases (Fig.~\ref{fig5}(b)). 
%
%
It is interesting to compare the relative contributions of vdW forces coming from the various 
vdW inclusive functionals.  
However, this not straightforward because the correlation energies in vdW$^{\rm TS}$ and the vdW-DFs contain
different terms (\emph{c.f.} Eqns.~\ref{eqn_TS} and ~\ref{eqn_vdw-df}).
Nonetheless, since in the vdW$^{\rm TS}$ scheme used here vdW is the correlation energy coming beyond GGA PBE ($E_c^{\rm LDA}+E_c^{\rm GGA}$)
we have computed a similar quantity from the vdW-DFs by subtracting $E_c^{\rm GGA}$ (with GGA PBE) from $E_c^{\rm NL}$.
The contribution of this modified non-local correlation energy to the lattice energy of ice is denoted as $\Delta E_c^\prime$
and is shown in Fig.~\ref{fig5}(c).
%
%
%
When we examine this term we find that it increases from the low to the high-density ice phases with all vdW inclusive functionals.
However, the magnitude of $\Delta E_c^\prime$ predicted by the different approaches differs significantly.
%
%
rPW86-vdW2 predicts the smallest $\Delta E_c^\prime$, revPBE-vdW and optPBE-vdW the largest,
and PBE/PBE0+vdW$^{\rm TS}$ falls in the middle.
%
%
Since a major component of $\Delta E_c^\prime$ is non-local vdW interactions  
the magnitude of $\Delta E_c^\prime$ should depend strongly on the vdW $C_6$ coefficients. 
Indeed, we find that the molecular $C_6$ coefficients calculated on ice (Table~\ref{tb4}) with the different functionals correlate well with
the relative magnitude of $\Delta E_c^\prime$.~\cite{note_on_c6_calculation}
Previous work showed that the $C_6$ coefficient of an isolated water molecule is $<$50\% too small with rPW86-vdW2 compared to experiment.~\cite{vydrov_pra_2010}
Here we find the same behavior for the $C_6$ coefficients of water molecules within all ice phases. 
Compared to all other vdW functionals the molecular $C_6$ coefficients obtained from rPW86-vdW2 are strikingly smaller, 40-50\% for ice I$h$ and 35-45\% for ice VIII.

To sum up, this brief analysis of the exchange and correlation contributions to the lattice energies has revealed that the large reduction 
in the exchange contribution to the lattice energy upon going from the low- to the high-density phases is compensated for by a growing correlation 
contribution to the lattice energy from the beyond GGA correlation ($\Delta E_c^\prime$). 
This compensation is obviously found for the vdW inclusive methods but not found for PBE and PBE0 and as a result the lattice energies 
are underestimated with PBE and PBE0 for the high pressure phases. 

\section{Conclusions}
\label{sec4} 


We have performed a detailed study on a selection of different ice phases with a range of~\emph{xc} functionals, 
including some of the recently developed functionals which account for vdW dispersion forces. 
Whilst we know a lot about the performance of these functionals in the gas phase 
(in particular on gas phase data sets such as the S22~\cite{s22,burns_jcp_11}) 
much less is known about how these functionals perform in the condensed phases, which was one of the key motivations for this study.
%
As seen before in the gas phase the vdW inclusive functionals do offer some improvement in performance.
This is particularly true for the relative energies of the different phases 
and as a result the phase transition pressures. 
However, the functionals tested are far from perfect and none simultaneously yields excellent lattice energies and
lattice constants for all phases. 
Of the schemes considered
PBE0+vdW$^{\rm TS}$ consistently overestimates lattice energies by $\sim$50 meV/H$_2$O and equilibrium densities by $\sim$5\%.
optPBE-vdW produces densities of ice that are in best agreement ($\sim$3\%) with experiment but the lattice energies are 
$\sim$50 meV/H$_2$O too large.
revPBE-vdW underestimates densities by $>$10\% and lattice energies by $\sim$50 meV/H$_2$O.
rPW86-vdW2 gives very accurate lattice energies but the densities are underestimated by $>$8\%.
%
%

The improved agreement between the experimental and calculated phase transition pressures when using the vdW functionals clearly highlights
the importance of accounting for vdW in ice.
%
However, even with vdW inclusive functionals, capturing all of the experimentally characterized ice phases 
on the water phase diagram is clearly still a major challenge and beyond the capabilities of the methods considered here.
Water is well known to provide a stern challenge for DFT, be it water clusters, liquid water and now ice.
The fact that several phases of ice are missing from the phase diagram of water is somewhat of a blow to the true
predictive ability of the methods considered here, but also a challenge and opportunity for developing and testing new methods.

From this study it is evident that the ice phases considered here are
extremely useful in providing a challenging ``data set'' against which new methods can be tested and proved.
It would of course be interesting to see how some of the other vdW inclusive DFT methods developed recently 
perform on the ice phases.
In this respect the already available experimental lattice energies and the matching DMC numbers are valuable references.
However, additional DMC data on other phases of ice would certainly be of value as would 
other vdW inclusive methods e.g., random-phase approximation~\cite{lu_prl_2009,schimka_nmat_2010} 
and  second order M\o ller-Plesset perturbation theory.~\cite{gruneis_jcp_2010,hermann_prl_08}
Finally, we note that the difficulty in predicting ice phases up to only the pressure range 1-2 GPa using GGA, hybrid, and vdW inclusive DFT approaches 
suggests that caution must be exercised when searching for and predicting new phases of water at yet higher pressures using such functionals.~\cite{pickard_2007,militzer_10,mcmahon_prb_2011,hermann_pnas_2012,pickard_prl_2013} \\

\textbf{Acknowledgements:} 
B. Santra and R.C. are supported by the Scientific Discovery through Advanced Computing (SciDAC) program 
funded by Department of Energy grant DE-SC0008626. 
A.M. was supported by the European Research Council (Quantum-CRASS project) and the Royal Society through a Royal Society Wolfson Research Merit Award. 
Also partly via our membership of the UK's HPC Materials Chemistry Consortium, which is funded by EPSRC (EP/F067496), 
this work made use of the computational facilities of HECToR.
J.K. is grateful to UCL and the EPSRC for support through the PhD+ scheme.
%
%


\begin{thebibliography}{132}
\expandafter\ifx\csname natexlab\endcsname\relax\def\natexlab#1{#1}\fi
\expandafter\ifx\csname bibnamefont\endcsname\relax
  \def\bibnamefont#1{#1}\fi
\expandafter\ifx\csname bibfnamefont\endcsname\relax
  \def\bibfnamefont#1{#1}\fi
\expandafter\ifx\csname citenamefont\endcsname\relax
  \def\citenamefont#1{#1}\fi
\expandafter\ifx\csname url\endcsname\relax
  \def\url#1{\texttt{#1}}\fi
\expandafter\ifx\csname urlprefix\endcsname\relax\def\urlprefix{URL }\fi
\providecommand{\bibinfo}[2]{#2}
\providecommand{\eprint}[2][]{\url{#2}}

\bibitem[{\citenamefont{Santra et~al.}(2007)\citenamefont{Santra, Michaelides,
  and Scheffler}}]{santra_jcp_2007}
\bibinfo{author}{\bibfnamefont{B.}~\bibnamefont{Santra}},
  \bibinfo{author}{\bibfnamefont{A.}~\bibnamefont{Michaelides}},
  \bibnamefont{and}
  \bibinfo{author}{\bibfnamefont{M.}~\bibnamefont{Scheffler}},
  \bibinfo{journal}{J. Chem. Phys.} \textbf{\bibinfo{volume}{127}},
  \bibinfo{pages}{184104} (\bibinfo{year}{2007}).

\bibitem[{\citenamefont{Santra et~al.}(2008)\citenamefont{Santra, Michaelides,
  Fuchs, Tkatchenko, Filippi, and Scheffler}}]{santra_jcp_2008}
\bibinfo{author}{\bibfnamefont{B.}~\bibnamefont{Santra}},
  \bibinfo{author}{\bibfnamefont{A.}~\bibnamefont{Michaelides}},
  \bibinfo{author}{\bibfnamefont{M.}~\bibnamefont{Fuchs}},
  \bibinfo{author}{\bibfnamefont{A.}~\bibnamefont{Tkatchenko}},
  \bibinfo{author}{\bibfnamefont{C.}~\bibnamefont{Filippi}}, \bibnamefont{and}
  \bibinfo{author}{\bibfnamefont{M.}~\bibnamefont{Scheffler}},
  \bibinfo{journal}{J. Chem. Phys.} \textbf{\bibinfo{volume}{129}},
  \bibinfo{pages}{194111} (\bibinfo{year}{2008}).

\bibitem[{\citenamefont{Santra et~al.}(2009)\citenamefont{Santra, Michaelides,
  and Scheffler}}]{santra_jcp_2009}
\bibinfo{author}{\bibfnamefont{B.}~\bibnamefont{Santra}},
  \bibinfo{author}{\bibfnamefont{A.}~\bibnamefont{Michaelides}},
  \bibnamefont{and}
  \bibinfo{author}{\bibfnamefont{M.}~\bibnamefont{Scheffler}},
  \bibinfo{journal}{J. Chem. Phys.} \textbf{\bibinfo{volume}{131}},
  \bibinfo{pages}{124509} (\bibinfo{year}{2009}).

\bibitem[{\citenamefont{Dahlke et~al.}(2008)\citenamefont{Dahlke, Olson,
  Leverentz, and Truhlar}}]{truhlar_hexamer_2008}
\bibinfo{author}{\bibfnamefont{E.~E.} \bibnamefont{Dahlke}},
  \bibinfo{author}{\bibfnamefont{R.~M.} \bibnamefont{Olson}},
  \bibinfo{author}{\bibfnamefont{H.~R.} \bibnamefont{Leverentz}},
  \bibnamefont{and} \bibinfo{author}{\bibfnamefont{D.~G.}
  \bibnamefont{Truhlar}}, \bibinfo{journal}{J. Phys. Chem. A}
  \textbf{\bibinfo{volume}{112}}, \bibinfo{pages}{3976} (\bibinfo{year}{2008}).

\bibitem[{\citenamefont{Xu and Goddard~III}(2004)}]{goddard_dimer_2004}
\bibinfo{author}{\bibfnamefont{X.}~\bibnamefont{Xu}} \bibnamefont{and}
  \bibinfo{author}{\bibfnamefont{W.~A.} \bibnamefont{Goddard~III}},
  \bibinfo{journal}{J. Phys. Chem. A} \textbf{\bibinfo{volume}{108}},
  \bibinfo{pages}{2305} (\bibinfo{year}{2004}).

\bibitem[{\citenamefont{Su et~al.}(2004)\citenamefont{Su, Xu, and
  Goddard~III}}]{goddard_hexamer_2004}
\bibinfo{author}{\bibfnamefont{J.~T.} \bibnamefont{Su}},
  \bibinfo{author}{\bibfnamefont{X.}~\bibnamefont{Xu}}, \bibnamefont{and}
  \bibinfo{author}{\bibfnamefont{W.~A.} \bibnamefont{Goddard~III}},
  \bibinfo{journal}{J. Phys. Chem. A} \textbf{\bibinfo{volume}{108}},
  \bibinfo{pages}{10518} (\bibinfo{year}{2004}).

\bibitem[{\citenamefont{Xantheas}(1995)}]{xantheas_1995}
\bibinfo{author}{\bibfnamefont{S.~S.} \bibnamefont{Xantheas}},
  \bibinfo{journal}{J. Chem. Phys.} \textbf{\bibinfo{volume}{102}},
  \bibinfo{pages}{4505} (\bibinfo{year}{1995}).

\bibitem[{\citenamefont{Kim and Jordan}(1994)}]{kim_jordan_1994}
\bibinfo{author}{\bibfnamefont{K.}~\bibnamefont{Kim}} \bibnamefont{and}
  \bibinfo{author}{\bibfnamefont{K.~D.} \bibnamefont{Jordan}},
  \bibinfo{journal}{J. Phys. Chem.} \textbf{\bibinfo{volume}{98}},
  \bibinfo{pages}{10089} (\bibinfo{year}{1994}).

\bibitem[{\citenamefont{Anderson and Tschumper}(2006)}]{tschumper_JCPA_2006}
\bibinfo{author}{\bibfnamefont{J.~A.} \bibnamefont{Anderson}} \bibnamefont{and}
  \bibinfo{author}{\bibfnamefont{G.~S.} \bibnamefont{Tschumper}},
  \bibinfo{journal}{J. Phys. Chem. A} \textbf{\bibinfo{volume}{110}},
  \bibinfo{pages}{7268} (\bibinfo{year}{2006}).

\bibitem[{\citenamefont{Shields and Kirschner}(2008)}]{shields_kirschner_2008}
\bibinfo{author}{\bibfnamefont{G.~C.} \bibnamefont{Shields}} \bibnamefont{and}
  \bibinfo{author}{\bibfnamefont{K.~N.} \bibnamefont{Kirschner}},
  \bibinfo{journal}{Synthesis and Reactivity in Inorganic, Metal-Organic, and
  Nano-Metal Chemistry} \textbf{\bibinfo{volume}{38}}, \bibinfo{pages}{32}
  (\bibinfo{year}{2008}).

\bibitem[{\citenamefont{Csonka et~al.}(2005)\citenamefont{Csonka, Ruzsinszky,
  and Perdew}}]{perdew_JCPB_2005}
\bibinfo{author}{\bibfnamefont{G.~I.} \bibnamefont{Csonka}},
  \bibinfo{author}{\bibfnamefont{A.}~\bibnamefont{Ruzsinszky}},
  \bibnamefont{and} \bibinfo{author}{\bibfnamefont{J.~P.}
  \bibnamefont{Perdew}}, \bibinfo{journal}{J. Phys. Chem. B}
  \textbf{\bibinfo{volume}{109}}, \bibinfo{pages}{21471}
  (\bibinfo{year}{2005}).

\bibitem[{\citenamefont{Dahlke and Truhlar}(2005)}]{truhlar_pbe1w_2005}
\bibinfo{author}{\bibfnamefont{E.~E.} \bibnamefont{Dahlke}} \bibnamefont{and}
  \bibinfo{author}{\bibfnamefont{D.~G.} \bibnamefont{Truhlar}},
  \bibinfo{journal}{J. Phys. Chem. B} \textbf{\bibinfo{volume}{109}},
  \bibinfo{pages}{15677} (\bibinfo{year}{2005}).

\bibitem[{\citenamefont{Gillan et~al.}(2012)\citenamefont{Gillan, Manby,
  Towler, and Alf\`e}}]{gillan_jcp_12}
\bibinfo{author}{\bibfnamefont{M.~J.} \bibnamefont{Gillan}},
  \bibinfo{author}{\bibfnamefont{F.~R.} \bibnamefont{Manby}},
  \bibinfo{author}{\bibfnamefont{M.~D.} \bibnamefont{Towler}},
  \bibnamefont{and} \bibinfo{author}{\bibfnamefont{D.}~\bibnamefont{Alf\`e}},
  \bibinfo{journal}{J. Chem. Phys.} \textbf{\bibinfo{volume}{136}},
  \bibinfo{pages}{244105} (\bibinfo{year}{2012}).

\bibitem[{\citenamefont{Ireta et~al.}(2004)\citenamefont{Ireta, Neugebauer, and
  Scheffler}}]{joel_pbe}
\bibinfo{author}{\bibfnamefont{J.}~\bibnamefont{Ireta}},
  \bibinfo{author}{\bibfnamefont{J.}~\bibnamefont{Neugebauer}},
  \bibnamefont{and}
  \bibinfo{author}{\bibfnamefont{M.}~\bibnamefont{Scheffler}},
  \bibinfo{journal}{J. Phys. Chem. A} \textbf{\bibinfo{volume}{108}},
  \bibinfo{pages}{5692} (\bibinfo{year}{2004}).

\bibitem[{\citenamefont{Tsuzuki and L\"uthi}(2001)}]{Tsuzuki}
\bibinfo{author}{\bibfnamefont{S.}~\bibnamefont{Tsuzuki}} \bibnamefont{and}
  \bibinfo{author}{\bibfnamefont{H.~P.} \bibnamefont{L\"uthi}},
  \bibinfo{journal}{J. Chem. Phys.} \textbf{\bibinfo{volume}{114}},
  \bibinfo{pages}{3949} (\bibinfo{year}{2001}).

\bibitem[{\citenamefont{Novoa and Sosa}(1995)}]{Novoa}
\bibinfo{author}{\bibfnamefont{J.~J.} \bibnamefont{Novoa}} \bibnamefont{and}
  \bibinfo{author}{\bibfnamefont{C.}~\bibnamefont{Sosa}}, \bibinfo{journal}{J.
  Phys. Chem.} \textbf{\bibinfo{volume}{99}}, \bibinfo{pages}{15837}
  (\bibinfo{year}{1995}).

\bibitem[{\citenamefont{Hammond et~al.}(2009)\citenamefont{Hammond, Govind,
  Kowalski, Autschbach, and Xantheas}}]{hammond_09}
\bibinfo{author}{\bibfnamefont{J.~R.} \bibnamefont{Hammond}},
  \bibinfo{author}{\bibfnamefont{N.}~\bibnamefont{Govind}},
  \bibinfo{author}{\bibfnamefont{K.}~\bibnamefont{Kowalski}},
  \bibinfo{author}{\bibfnamefont{J.}~\bibnamefont{Autschbach}},
  \bibnamefont{and} \bibinfo{author}{\bibfnamefont{S.~S.}
  \bibnamefont{Xantheas}}, \bibinfo{journal}{J. Chem. Phys.}
  \textbf{\bibinfo{volume}{131}}, \bibinfo{pages}{214103}
  (\bibinfo{year}{2009}).

\bibitem[{\citenamefont{Wang et~al.}(2010)\citenamefont{Wang, Jenness,
  Al-Saidi, and Jordan}}]{jordan_10}
\bibinfo{author}{\bibfnamefont{F.-F.} \bibnamefont{Wang}},
  \bibinfo{author}{\bibfnamefont{G.}~\bibnamefont{Jenness}},
  \bibinfo{author}{\bibfnamefont{W.~A.} \bibnamefont{Al-Saidi}},
  \bibnamefont{and} \bibinfo{author}{\bibfnamefont{K.~D.}
  \bibnamefont{Jordan}}, \bibinfo{journal}{J. Chem. Phys.}
  \textbf{\bibinfo{volume}{132}}, \bibinfo{pages}{134303}
  (\bibinfo{year}{2010}).

\bibitem[{\citenamefont{Grossman et~al.}(2004)\citenamefont{Grossman,
  Schwegler, Draeger, Gygi, and Galli}}]{grossman_water_1}
\bibinfo{author}{\bibfnamefont{J.~C.} \bibnamefont{Grossman}},
  \bibinfo{author}{\bibfnamefont{E.}~\bibnamefont{Schwegler}},
  \bibinfo{author}{\bibfnamefont{E.~W.} \bibnamefont{Draeger}},
  \bibinfo{author}{\bibfnamefont{F.}~\bibnamefont{Gygi}}, \bibnamefont{and}
  \bibinfo{author}{\bibfnamefont{G.}~\bibnamefont{Galli}}, \bibinfo{journal}{J.
  Chem. Phys.} \textbf{\bibinfo{volume}{120}}, \bibinfo{pages}{300}
  (\bibinfo{year}{2004}).

\bibitem[{\citenamefont{Schwegler et~al.}(2004)\citenamefont{Schwegler,
  Grossman, Gygi, and Galli}}]{grossman_water_2}
\bibinfo{author}{\bibfnamefont{E.}~\bibnamefont{Schwegler}},
  \bibinfo{author}{\bibfnamefont{J.~C.} \bibnamefont{Grossman}},
  \bibinfo{author}{\bibfnamefont{F.}~\bibnamefont{Gygi}}, \bibnamefont{and}
  \bibinfo{author}{\bibfnamefont{G.}~\bibnamefont{Galli}}, \bibinfo{journal}{J.
  Chem. Phys.} \textbf{\bibinfo{volume}{121}}, \bibinfo{pages}{5400}
  (\bibinfo{year}{2004}).

\bibitem[{\citenamefont{Fern\'{a}ndez-Serra and Artacho}(2004)}]{artacho}
\bibinfo{author}{\bibfnamefont{M.~V.} \bibnamefont{Fern\'{a}ndez-Serra}}
  \bibnamefont{and} \bibinfo{author}{\bibfnamefont{E.}~\bibnamefont{Artacho}},
  \bibinfo{journal}{J. Chem. Phys.} \textbf{\bibinfo{volume}{121}},
  \bibinfo{pages}{11136} (\bibinfo{year}{2004}).

\bibitem[{\citenamefont{McGrath et~al.}(2005)\citenamefont{McGrath, Siepmann,
  Kuo, Mundy, VandeVondele, Hutter, Mohamed, and Krack}}]{McGrath_05}
\bibinfo{author}{\bibfnamefont{M.~J.} \bibnamefont{McGrath}},
  \bibinfo{author}{\bibfnamefont{J.~I.} \bibnamefont{Siepmann}},
  \bibinfo{author}{\bibfnamefont{I.-F.~W.} \bibnamefont{Kuo}},
  \bibinfo{author}{\bibfnamefont{C.~J.} \bibnamefont{Mundy}},
  \bibinfo{author}{\bibfnamefont{J.}~\bibnamefont{VandeVondele}},
  \bibinfo{author}{\bibfnamefont{J.}~\bibnamefont{Hutter}},
  \bibinfo{author}{\bibfnamefont{F.}~\bibnamefont{Mohamed}}, \bibnamefont{and}
  \bibinfo{author}{\bibfnamefont{M.}~\bibnamefont{Krack}},
  \bibinfo{journal}{ChemPhysChem} \textbf{\bibinfo{volume}{6}},
  \bibinfo{pages}{1894} (\bibinfo{year}{2005}).

\bibitem[{\citenamefont{McGrath
  et~al.}(2006{\natexlab{a}})\citenamefont{McGrath, Siepmann, Kuo, and
  Mundy}}]{McGrath_06}
\bibinfo{author}{\bibfnamefont{M.~J.} \bibnamefont{McGrath}},
  \bibinfo{author}{\bibfnamefont{J.~I.} \bibnamefont{Siepmann}},
  \bibinfo{author}{\bibfnamefont{I.-F.~W.} \bibnamefont{Kuo}},
  \bibnamefont{and} \bibinfo{author}{\bibfnamefont{C.}~\bibnamefont{Mundy}},
  \bibinfo{journal}{Molec. Phys.} \textbf{\bibinfo{volume}{104}},
  \bibinfo{pages}{3619} (\bibinfo{year}{2006}{\natexlab{a}}).

\bibitem[{\citenamefont{McGrath
  et~al.}(2006{\natexlab{b}})\citenamefont{McGrath, Siepmann, Kuo, Mundy,
  VandeVondele, Hutter, Mohamed, and Krack}}]{Mcgrath_jpcA_06}
\bibinfo{author}{\bibfnamefont{M.~J.} \bibnamefont{McGrath}},
  \bibinfo{author}{\bibfnamefont{J.~I.} \bibnamefont{Siepmann}},
  \bibinfo{author}{\bibfnamefont{I.-F.~W.} \bibnamefont{Kuo}},
  \bibinfo{author}{\bibfnamefont{C.~J.} \bibnamefont{Mundy}},
  \bibinfo{author}{\bibfnamefont{J.}~\bibnamefont{VandeVondele}},
  \bibinfo{author}{\bibfnamefont{J.}~\bibnamefont{Hutter}},
  \bibinfo{author}{\bibfnamefont{F.}~\bibnamefont{Mohamed}}, \bibnamefont{and}
  \bibinfo{author}{\bibfnamefont{M.}~\bibnamefont{Krack}}, \bibinfo{journal}{J.
  Phys. Chem. A} \textbf{\bibinfo{volume}{110}}, \bibinfo{pages}{640}
  (\bibinfo{year}{2006}{\natexlab{b}}).

\bibitem[{\citenamefont{Lee and Tuckerman}(2006)}]{tuckerman_jcp_06}
\bibinfo{author}{\bibfnamefont{H.-S.} \bibnamefont{Lee}} \bibnamefont{and}
  \bibinfo{author}{\bibfnamefont{M.~E.} \bibnamefont{Tuckerman}},
  \bibinfo{journal}{J. Chem. Phys.} \textbf{\bibinfo{volume}{125}},
  \bibinfo{pages}{154507} (\bibinfo{year}{2006}).

\bibitem[{\citenamefont{Lee and Tuckerman}(2007)}]{tuckerman_jcp_07}
\bibinfo{author}{\bibfnamefont{H.-S.} \bibnamefont{Lee}} \bibnamefont{and}
  \bibinfo{author}{\bibfnamefont{M.~E.} \bibnamefont{Tuckerman}},
  \bibinfo{journal}{J. Chem. Phys.} \textbf{\bibinfo{volume}{126}},
  \bibinfo{pages}{164501} (\bibinfo{year}{2007}).

\bibitem[{\citenamefont{Todorova et~al.}(2006)\citenamefont{Todorova,
  Seitsonen, Hutter, Kuo, and Mundy}}]{Todorova}
\bibinfo{author}{\bibfnamefont{T.}~\bibnamefont{Todorova}},
  \bibinfo{author}{\bibfnamefont{A.~P.} \bibnamefont{Seitsonen}},
  \bibinfo{author}{\bibfnamefont{J.}~\bibnamefont{Hutter}},
  \bibinfo{author}{\bibfnamefont{I.-F.~W.} \bibnamefont{Kuo}},
  \bibnamefont{and} \bibinfo{author}{\bibfnamefont{C.~J.} \bibnamefont{Mundy}},
  \bibinfo{journal}{J. Phys. Chem. B} \textbf{\bibinfo{volume}{110}},
  \bibinfo{pages}{3685} (\bibinfo{year}{2006}).

\bibitem[{\citenamefont{VandeVondele et~al.}(2005)\citenamefont{VandeVondele,
  Mohamed, Krack, Hutter, Sprik, and Parrinello}}]{vandevondele_05}
\bibinfo{author}{\bibfnamefont{J.}~\bibnamefont{VandeVondele}},
  \bibinfo{author}{\bibfnamefont{F.}~\bibnamefont{Mohamed}},
  \bibinfo{author}{\bibfnamefont{M.}~\bibnamefont{Krack}},
  \bibinfo{author}{\bibfnamefont{J.}~\bibnamefont{Hutter}},
  \bibinfo{author}{\bibfnamefont{M.}~\bibnamefont{Sprik}}, \bibnamefont{and}
  \bibinfo{author}{\bibfnamefont{M.}~\bibnamefont{Parrinello}},
  \bibinfo{journal}{J. Chem. Phys.} \textbf{\bibinfo{volume}{122}},
  \bibinfo{pages}{014515} (\bibinfo{year}{2005}).

\bibitem[{\citenamefont{Guidon et~al.}(2008)\citenamefont{Guidon, Schiffmann,
  Hutter, and VandeVondele}}]{vandevondele_08}
\bibinfo{author}{\bibfnamefont{M.}~\bibnamefont{Guidon}},
  \bibinfo{author}{\bibfnamefont{F.}~\bibnamefont{Schiffmann}},
  \bibinfo{author}{\bibfnamefont{J.}~\bibnamefont{Hutter}}, \bibnamefont{and}
  \bibinfo{author}{\bibfnamefont{J.}~\bibnamefont{VandeVondele}},
  \bibinfo{journal}{J. Chem. Phys.} \textbf{\bibinfo{volume}{128}},
  \bibinfo{pages}{214104} (\bibinfo{year}{2008}).

\bibitem[{\citenamefont{K\"{u}hne et~al.}(2009)\citenamefont{K\"{u}hne, Krack,
  and Parrinello}}]{parrinello_09}
\bibinfo{author}{\bibfnamefont{T.~D.} \bibnamefont{K\"{u}hne}},
  \bibinfo{author}{\bibfnamefont{M.}~\bibnamefont{Krack}}, \bibnamefont{and}
  \bibinfo{author}{\bibfnamefont{M.}~\bibnamefont{Parrinello}},
  \bibinfo{journal}{J. Chem. Theory Comput.} \textbf{\bibinfo{volume}{5}},
  \bibinfo{pages}{235} (\bibinfo{year}{2009}).

\bibitem[{\citenamefont{Sit and Marzari}(2005)}]{sit_marzari_jcp_05}
\bibinfo{author}{\bibfnamefont{P.~H.-L.} \bibnamefont{Sit}} \bibnamefont{and}
  \bibinfo{author}{\bibfnamefont{N.}~\bibnamefont{Marzari}},
  \bibinfo{journal}{J. Chem. Phys.} \textbf{\bibinfo{volume}{122}},
  \bibinfo{pages}{204510} (\bibinfo{year}{2005}).

\bibitem[{\citenamefont{Fern\'{a}ndez-Serra
  et~al.}(2005)\citenamefont{Fern\'{a}ndez-Serra, Ferlat, and
  Artacho}}]{artacho_05}
\bibinfo{author}{\bibfnamefont{M.~V.} \bibnamefont{Fern\'{a}ndez-Serra}},
  \bibinfo{author}{\bibfnamefont{G.}~\bibnamefont{Ferlat}}, \bibnamefont{and}
  \bibinfo{author}{\bibfnamefont{E.}~\bibnamefont{Artacho}},
  \bibinfo{journal}{Molecular Simualtion} \textbf{\bibinfo{volume}{31}},
  \bibinfo{pages}{361} (\bibinfo{year}{2005}).

\bibitem[{\citenamefont{Asthagiri et~al.}(2003)\citenamefont{Asthagiri, Pratt,
  and Kress}}]{asthagiri_03}
\bibinfo{author}{\bibfnamefont{D.}~\bibnamefont{Asthagiri}},
  \bibinfo{author}{\bibfnamefont{L.~R.} \bibnamefont{Pratt}}, \bibnamefont{and}
  \bibinfo{author}{\bibfnamefont{J.~D.} \bibnamefont{Kress}},
  \bibinfo{journal}{Phys. Rev. E} \textbf{\bibinfo{volume}{68}},
  \bibinfo{pages}{041505} (\bibinfo{year}{2003}).

\bibitem[{\citenamefont{Mattson and Mattson}(2009)}]{am05_water_09}
\bibinfo{author}{\bibfnamefont{A.~E.} \bibnamefont{Mattson}} \bibnamefont{and}
  \bibinfo{author}{\bibfnamefont{T.~R.} \bibnamefont{Mattson}},
  \bibinfo{journal}{J. Chem. Theory Comput.} \textbf{\bibinfo{volume}{5}},
  \bibinfo{pages}{887} (\bibinfo{year}{2009}).

\bibitem[{\citenamefont{Kuo et~al.}(2004)\citenamefont{Kuo, Mundy, McGrath,
  Siepmann, VandeVondele, Sprik, Hutter, Chen, Klein, Mohamed et~al.}}]{Kuo_04}
\bibinfo{author}{\bibfnamefont{I.-F.~W.} \bibnamefont{Kuo}},
  \bibinfo{author}{\bibfnamefont{C.~J.} \bibnamefont{Mundy}},
  \bibinfo{author}{\bibfnamefont{M.~J.} \bibnamefont{McGrath}},
  \bibinfo{author}{\bibfnamefont{J.~I.} \bibnamefont{Siepmann}},
  \bibinfo{author}{\bibfnamefont{J.}~\bibnamefont{VandeVondele}},
  \bibinfo{author}{\bibfnamefont{M.}~\bibnamefont{Sprik}},
  \bibinfo{author}{\bibfnamefont{J.}~\bibnamefont{Hutter}},
  \bibinfo{author}{\bibfnamefont{B.}~\bibnamefont{Chen}},
  \bibinfo{author}{\bibfnamefont{M.~L.} \bibnamefont{Klein}},
  \bibinfo{author}{\bibfnamefont{F.}~\bibnamefont{Mohamed}},
  \bibnamefont{et~al.}, \bibinfo{journal}{J. Phys. Chem. B}
  \textbf{\bibinfo{volume}{108}}, \bibinfo{pages}{12990}
  (\bibinfo{year}{2004}).

\bibitem[{\citenamefont{Sprik et~al.}(1996)\citenamefont{Sprik, Hutter, and
  Parrinello}}]{sprik_jcp_96}
\bibinfo{author}{\bibfnamefont{M.}~\bibnamefont{Sprik}},
  \bibinfo{author}{\bibfnamefont{J.}~\bibnamefont{Hutter}}, \bibnamefont{and}
  \bibinfo{author}{\bibfnamefont{M.}~\bibnamefont{Parrinello}},
  \bibinfo{journal}{J. Chem. Phys.} \textbf{\bibinfo{volume}{105}},
  \bibinfo{pages}{1142} (\bibinfo{year}{1996}).

\bibitem[{\citenamefont{Silvestrelli and
  Parrinello}(1999)}]{silvestrlli_jcp_99}
\bibinfo{author}{\bibfnamefont{P.~L.} \bibnamefont{Silvestrelli}}
  \bibnamefont{and}
  \bibinfo{author}{\bibfnamefont{M.}~\bibnamefont{Parrinello}},
  \bibinfo{journal}{J. Chem. Phys.} \textbf{\bibinfo{volume}{111}},
  \bibinfo{pages}{3572} (\bibinfo{year}{1999}).

\bibitem[{\citenamefont{Yoo et~al.}(2009)\citenamefont{Yoo, Zeng, and
  Xantheas}}]{xantheas_JCP_09}
\bibinfo{author}{\bibfnamefont{S.}~\bibnamefont{Yoo}},
  \bibinfo{author}{\bibfnamefont{X.~C.} \bibnamefont{Zeng}}, \bibnamefont{and}
  \bibinfo{author}{\bibfnamefont{S.~S.} \bibnamefont{Xantheas}},
  \bibinfo{journal}{J. Chem. Phys.} \textbf{\bibinfo{volume}{130}},
  \bibinfo{pages}{221102} (\bibinfo{year}{2009}).

\bibitem[{\citenamefont{Zhang et~al.}(2011{\natexlab{a}})\citenamefont{Zhang,
  Donadio, Gygi, and Galli}}]{zhang_jctc_2011_2}
\bibinfo{author}{\bibfnamefont{C.}~\bibnamefont{Zhang}},
  \bibinfo{author}{\bibfnamefont{D.}~\bibnamefont{Donadio}},
  \bibinfo{author}{\bibfnamefont{F.}~\bibnamefont{Gygi}}, \bibnamefont{and}
  \bibinfo{author}{\bibfnamefont{G.}~\bibnamefont{Galli}}, \bibinfo{journal}{J.
  Chem. Theory Comput.} \textbf{\bibinfo{volume}{7}}, \bibinfo{pages}{1443}
  (\bibinfo{year}{2011}{\natexlab{a}}).

\bibitem[{\citenamefont{Chen et~al.}(2003)\citenamefont{Chen, Ivanov, Klein,
  and Parrinello}}]{parrinello_03}
\bibinfo{author}{\bibfnamefont{B.}~\bibnamefont{Chen}},
  \bibinfo{author}{\bibfnamefont{I.}~\bibnamefont{Ivanov}},
  \bibinfo{author}{\bibfnamefont{M.~L.} \bibnamefont{Klein}}, \bibnamefont{and}
  \bibinfo{author}{\bibfnamefont{M.}~\bibnamefont{Parrinello}},
  \bibinfo{journal}{Phys. Rev. Lett.} \textbf{\bibinfo{volume}{91}},
  \bibinfo{pages}{215503} (\bibinfo{year}{2003}).

\bibitem[{\citenamefont{Morrone and Car}(2008)}]{car_08}
\bibinfo{author}{\bibfnamefont{J.~A.} \bibnamefont{Morrone}} \bibnamefont{and}
  \bibinfo{author}{\bibfnamefont{R.}~\bibnamefont{Car}},
  \bibinfo{journal}{Phys. Rev. Lett.} \textbf{\bibinfo{volume}{101}},
  \bibinfo{pages}{017801} (\bibinfo{year}{2008}).

\bibitem[{\citenamefont{Laasonen et~al.}(1993)\citenamefont{Laasonen, Sprik,
  Parrinello, and Car}}]{DFT_MD_93}
\bibinfo{author}{\bibfnamefont{K.}~\bibnamefont{Laasonen}},
  \bibinfo{author}{\bibfnamefont{M.}~\bibnamefont{Sprik}},
  \bibinfo{author}{\bibfnamefont{M.}~\bibnamefont{Parrinello}},
  \bibnamefont{and} \bibinfo{author}{\bibfnamefont{R.}~\bibnamefont{Car}},
  \bibinfo{journal}{J. Chem. Phys.} \textbf{\bibinfo{volume}{99}},
  \bibinfo{pages}{9080} (\bibinfo{year}{1993}).

\bibitem[{\citenamefont{Akin-Ojo and Wang}(2011)}]{ojo_cpl_2011}
\bibinfo{author}{\bibfnamefont{O.}~\bibnamefont{Akin-Ojo}} \bibnamefont{and}
  \bibinfo{author}{\bibfnamefont{F.}~\bibnamefont{Wang}},
  \bibinfo{journal}{Chem. Phys. Lett.} \textbf{\bibinfo{volume}{513}},
  \bibinfo{pages}{59} (\bibinfo{year}{2011}).

\bibitem[{\citenamefont{Alf\`e et~al.}(2013)\citenamefont{Alf\`e, Bart\'ok,
  Cs\'anyi, and Gillan}}]{gillan_jcp_13}
\bibinfo{author}{\bibfnamefont{D.}~\bibnamefont{Alf\`e}},
  \bibinfo{author}{\bibfnamefont{A.~P.} \bibnamefont{Bart\'ok}},
  \bibinfo{author}{\bibfnamefont{G.}~\bibnamefont{Cs\'anyi}}, \bibnamefont{and}
  \bibinfo{author}{\bibfnamefont{M.~J.} \bibnamefont{Gillan}},
  \bibinfo{journal}{J. Chem. Phys.} \textbf{\bibinfo{volume}{138}},
  \bibinfo{pages}{221102} (\bibinfo{year}{2013}).

\bibitem[{\citenamefont{Santra et~al.}(2011)\citenamefont{Santra, Klime\v{s},
  Alf\`e, Tkatchenko, Slater, Michaelides, Car, and
  Scheffler}}]{santra_prl_2011}
\bibinfo{author}{\bibfnamefont{B.}~\bibnamefont{Santra}},
  \bibinfo{author}{\bibfnamefont{J.}~\bibnamefont{Klime\v{s}}},
  \bibinfo{author}{\bibfnamefont{D.}~\bibnamefont{Alf\`e}},
  \bibinfo{author}{\bibfnamefont{A.}~\bibnamefont{Tkatchenko}},
  \bibinfo{author}{\bibfnamefont{B.}~\bibnamefont{Slater}},
  \bibinfo{author}{\bibfnamefont{A.}~\bibnamefont{Michaelides}},
  \bibinfo{author}{\bibfnamefont{R.}~\bibnamefont{Car}}, \bibnamefont{and}
  \bibinfo{author}{\bibfnamefont{M.}~\bibnamefont{Scheffler}},
  \bibinfo{journal}{Phys. Rev. Lett.} \textbf{\bibinfo{volume}{107}},
  \bibinfo{pages}{185701} (\bibinfo{year}{2011}).

\bibitem[{\citenamefont{Murray and Galli}(2012)}]{murray_prl_2012}
\bibinfo{author}{\bibfnamefont{E.~D.} \bibnamefont{Murray}} \bibnamefont{and}
  \bibinfo{author}{\bibfnamefont{G.}~\bibnamefont{Galli}},
  \bibinfo{journal}{Phys. Rev. Lett.} \textbf{\bibinfo{volume}{108}},
  \bibinfo{pages}{105502} (\bibinfo{year}{2012}).

\bibitem[{\citenamefont{Hamada}(2010)}]{hamada_10}
\bibinfo{author}{\bibfnamefont{I.}~\bibnamefont{Hamada}}, \bibinfo{journal}{J.
  Chem. Phys.} \textbf{\bibinfo{volume}{133}}, \bibinfo{pages}{214503}
  (\bibinfo{year}{2010}).

\bibitem[{\citenamefont{Kolb and Thonhauser}(2011)}]{kolb_prb_11}
\bibinfo{author}{\bibfnamefont{B.}~\bibnamefont{Kolb}} \bibnamefont{and}
  \bibinfo{author}{\bibfnamefont{T.}~\bibnamefont{Thonhauser}},
  \bibinfo{journal}{Phys. Rev. B} \textbf{\bibinfo{volume}{84}},
  \bibinfo{pages}{045116} (\bibinfo{year}{2011}).

\bibitem[{\citenamefont{Kambara et~al.}(2012)\citenamefont{Kambara, Takahashi,
  Hayashi, and Kuo}}]{pccp_ice_12}
\bibinfo{author}{\bibfnamefont{O.}~\bibnamefont{Kambara}},
  \bibinfo{author}{\bibfnamefont{K.}~\bibnamefont{Takahashi}},
  \bibinfo{author}{\bibfnamefont{M.}~\bibnamefont{Hayashi}}, \bibnamefont{and}
  \bibinfo{author}{\bibfnamefont{J.-L.} \bibnamefont{Kuo}},
  \bibinfo{journal}{Phys. Chem. Chem. Phys.} \textbf{\bibinfo{volume}{14}},
  \bibinfo{pages}{11484} (\bibinfo{year}{2012}).

\bibitem[{\citenamefont{Fang et~al.}(2013)\citenamefont{Fang, Xiao, Tao, Sun,
  and Perdew}}]{fang_prb_2013}
\bibinfo{author}{\bibfnamefont{Y.}~\bibnamefont{Fang}},
  \bibinfo{author}{\bibfnamefont{B.}~\bibnamefont{Xiao}},
  \bibinfo{author}{\bibfnamefont{J.}~\bibnamefont{Tao}},
  \bibinfo{author}{\bibfnamefont{J.}~\bibnamefont{Sun}}, \bibnamefont{and}
  \bibinfo{author}{\bibfnamefont{J.~P.} \bibnamefont{Perdew}},
  \bibinfo{journal}{Phys. Rev. B} \textbf{\bibinfo{volume}{87}},
  \bibinfo{pages}{214101} (\bibinfo{year}{2013}).

\bibitem[{\citenamefont{Gillan et~al.}(2013)\citenamefont{Gillan, Alf\`e,
  Bart\'ok, and Cs\'anyi}}]{gillan_arxiv_1303.0751}
\bibinfo{author}{\bibfnamefont{M.~J.} \bibnamefont{Gillan}},
  \bibinfo{author}{\bibfnamefont{D.}~\bibnamefont{Alf\`e}},
  \bibinfo{author}{\bibfnamefont{A.~P.} \bibnamefont{Bart\'ok}},
  \bibnamefont{and} \bibinfo{author}{\bibfnamefont{G.}~\bibnamefont{Cs\'anyi}},
  \bibinfo{journal}{arXiv:1303.0751}  (\bibinfo{year}{2013}),
  \url{http://arxiv.org/abs/1303.0751}.

\bibitem[{\citenamefont{Pamuk et~al.}(2012)\citenamefont{Pamuk, Soler,
  Ram\'irez, Herrero, Stephens, Allen, and Fern\'andez-Serra}}]{pamuk_prl_2012}
\bibinfo{author}{\bibfnamefont{B.}~\bibnamefont{Pamuk}},
  \bibinfo{author}{\bibfnamefont{J.~M.} \bibnamefont{Soler}},
  \bibinfo{author}{\bibfnamefont{R.}~\bibnamefont{Ram\'irez}},
  \bibinfo{author}{\bibfnamefont{C.~P.} \bibnamefont{Herrero}},
  \bibinfo{author}{\bibfnamefont{P.}~\bibnamefont{Stephens}},
  \bibinfo{author}{\bibfnamefont{P.~B.} \bibnamefont{Allen}}, \bibnamefont{and}
  \bibinfo{author}{\bibfnamefont{M.-V.} \bibnamefont{Fern\'andez-Serra}},
  \bibinfo{journal}{Phys. Rev. Lett.} \textbf{\bibinfo{volume}{108}},
  \bibinfo{pages}{193003} (\bibinfo{year}{2012}).

\bibitem[{\citenamefont{Lee et~al.}(1992)\citenamefont{Lee, Vanderbilt,
  Laasonen, Car, and Parrinello}}]{car_ice_92}
\bibinfo{author}{\bibfnamefont{C.}~\bibnamefont{Lee}},
  \bibinfo{author}{\bibfnamefont{D.}~\bibnamefont{Vanderbilt}},
  \bibinfo{author}{\bibfnamefont{K.}~\bibnamefont{Laasonen}},
  \bibinfo{author}{\bibfnamefont{R.}~\bibnamefont{Car}}, \bibnamefont{and}
  \bibinfo{author}{\bibfnamefont{M.}~\bibnamefont{Parrinello}},
  \bibinfo{journal}{Phys. Rev. Lett.} \textbf{\bibinfo{volume}{69}},
  \bibinfo{pages}{462} (\bibinfo{year}{1992}).

\bibitem[{\citenamefont{Hamann}(1997)}]{hamann_97}
\bibinfo{author}{\bibfnamefont{D.~R.} \bibnamefont{Hamann}},
  \bibinfo{journal}{Phys. Rev. B} \textbf{\bibinfo{volume}{55}},
  \bibinfo{pages}{R10157} (\bibinfo{year}{1997}).

\bibitem[{\citenamefont{Singer et~al.}(2005)\citenamefont{Singer, Kuo, Hirsch,
  Knight, Ojam{\"a}e, and Klein}}]{klein_05}
\bibinfo{author}{\bibfnamefont{S.~J.} \bibnamefont{Singer}},
  \bibinfo{author}{\bibfnamefont{J.-L.} \bibnamefont{Kuo}},
  \bibinfo{author}{\bibfnamefont{T.~K.} \bibnamefont{Hirsch}},
  \bibinfo{author}{\bibfnamefont{C.}~\bibnamefont{Knight}},
  \bibinfo{author}{\bibfnamefont{L.}~\bibnamefont{Ojam{\"a}e}},
  \bibnamefont{and} \bibinfo{author}{\bibfnamefont{M.~L.} \bibnamefont{Klein}},
  \bibinfo{journal}{Phys. Rev. Lett.} \textbf{\bibinfo{volume}{94}},
  \bibinfo{pages}{135701} (\bibinfo{year}{2005}).

\bibitem[{\citenamefont{Tribello et~al.}(2006)\citenamefont{Tribello, Slater,
  and Salzmann}}]{slater_jacs_06}
\bibinfo{author}{\bibfnamefont{G.~A.} \bibnamefont{Tribello}},
  \bibinfo{author}{\bibfnamefont{B.}~\bibnamefont{Slater}}, \bibnamefont{and}
  \bibinfo{author}{\bibfnamefont{C.~G.} \bibnamefont{Salzmann}},
  \bibinfo{journal}{J. Am. Chem. Soc.} \textbf{\bibinfo{volume}{128}},
  \bibinfo{pages}{12594} (\bibinfo{year}{2006}).

\bibitem[{\citenamefont{{de Koning} et~al.}(2006)\citenamefont{{de Koning},
  Antonelli, {da Silva}, and Fazzio}}]{de_koning_06}
\bibinfo{author}{\bibfnamefont{M.}~\bibnamefont{{de Koning}}},
  \bibinfo{author}{\bibfnamefont{A.}~\bibnamefont{Antonelli}},
  \bibinfo{author}{\bibfnamefont{A.~J.~R.} \bibnamefont{{da Silva}}},
  \bibnamefont{and} \bibinfo{author}{\bibfnamefont{A.}~\bibnamefont{Fazzio}},
  \bibinfo{journal}{Phys. Rev. Lett.} \textbf{\bibinfo{volume}{96}},
  \bibinfo{pages}{075501} (\bibinfo{year}{2006}).

\bibitem[{\citenamefont{Feibelman}(2008)}]{feibelman_08}
\bibinfo{author}{\bibfnamefont{P.~J.} \bibnamefont{Feibelman}},
  \bibinfo{journal}{Phys. Chem. Chem. Phys.} \textbf{\bibinfo{volume}{10}},
  \bibinfo{pages}{4688} (\bibinfo{year}{2008}).

\bibitem[{\citenamefont{Hermann and Schwerdtfeger}(2008)}]{hermann_prl_08}
\bibinfo{author}{\bibfnamefont{A.}~\bibnamefont{Hermann}} \bibnamefont{and}
  \bibinfo{author}{\bibfnamefont{P.}~\bibnamefont{Schwerdtfeger}},
  \bibinfo{journal}{Phys. Rev. Lett.} \textbf{\bibinfo{volume}{101}},
  \bibinfo{pages}{183005} (\bibinfo{year}{2008}).

\bibitem[{\citenamefont{Pan et~al.}(2008)\citenamefont{Pan, Liu, Tribello,
  Slater, Michaelides, and Wang}}]{pan_08}
\bibinfo{author}{\bibfnamefont{D.}~\bibnamefont{Pan}},
  \bibinfo{author}{\bibfnamefont{L.-M.} \bibnamefont{Liu}},
  \bibinfo{author}{\bibfnamefont{G.~A.} \bibnamefont{Tribello}},
  \bibinfo{author}{\bibfnamefont{B.}~\bibnamefont{Slater}},
  \bibinfo{author}{\bibfnamefont{A.}~\bibnamefont{Michaelides}},
  \bibnamefont{and} \bibinfo{author}{\bibfnamefont{E.}~\bibnamefont{Wang}},
  \bibinfo{journal}{Phys. Rev. Lett.} \textbf{\bibinfo{volume}{101}},
  \bibinfo{pages}{155703} (\bibinfo{year}{2008}).

\bibitem[{\citenamefont{Erba et~al.}(2009)\citenamefont{Erba, Casassa, Maschio,
  and Pisani}}]{pisani_09}
\bibinfo{author}{\bibfnamefont{A.}~\bibnamefont{Erba}},
  \bibinfo{author}{\bibfnamefont{S.}~\bibnamefont{Casassa}},
  \bibinfo{author}{\bibfnamefont{L.}~\bibnamefont{Maschio}}, \bibnamefont{and}
  \bibinfo{author}{\bibfnamefont{C.}~\bibnamefont{Pisani}},
  \bibinfo{journal}{J. Phys. Chem. B} \textbf{\bibinfo{volume}{113}},
  \bibinfo{pages}{2347} (\bibinfo{year}{2009}).

\bibitem[{\citenamefont{Morrone et~al.}(2009)\citenamefont{Morrone, Lin, and
  Car}}]{car_09}
\bibinfo{author}{\bibfnamefont{J.~A.} \bibnamefont{Morrone}},
  \bibinfo{author}{\bibfnamefont{L.}~\bibnamefont{Lin}}, \bibnamefont{and}
  \bibinfo{author}{\bibfnamefont{R.}~\bibnamefont{Car}}, \bibinfo{journal}{J.
  Chem. Phys.} \textbf{\bibinfo{volume}{130}}, \bibinfo{pages}{204511}
  (\bibinfo{year}{2009}).

\bibitem[{\citenamefont{Militzer and Wilson}(2010)}]{militzer_10}
\bibinfo{author}{\bibfnamefont{B.}~\bibnamefont{Militzer}} \bibnamefont{and}
  \bibinfo{author}{\bibfnamefont{H.~F.} \bibnamefont{Wilson}},
  \bibinfo{journal}{Phys. Rev. Lett.} \textbf{\bibinfo{volume}{105}},
  \bibinfo{pages}{195701} (\bibinfo{year}{2010}).

\bibitem[{\citenamefont{Labat et~al.}(2011)\citenamefont{Labat, Pouchan, Adamo,
  and Scuseria}}]{labat_11}
\bibinfo{author}{\bibfnamefont{F.}~\bibnamefont{Labat}},
  \bibinfo{author}{\bibfnamefont{C.}~\bibnamefont{Pouchan}},
  \bibinfo{author}{\bibfnamefont{C.}~\bibnamefont{Adamo}}, \bibnamefont{and}
  \bibinfo{author}{\bibfnamefont{G.~E.} \bibnamefont{Scuseria}},
  \bibinfo{journal}{J. Comput. Chem.} \textbf{\bibinfo{volume}{32}},
  \bibinfo{pages}{2177} (\bibinfo{year}{2011}).

\bibitem[{\citenamefont{Lin et~al.}(2009)\citenamefont{Lin, Seitsonen,
  Coutinho-Neto, Tavernelli, and Rothlisberger}}]{lin_jpcB_09}
\bibinfo{author}{\bibfnamefont{I.-C.} \bibnamefont{Lin}},
  \bibinfo{author}{\bibfnamefont{A.~P.} \bibnamefont{Seitsonen}},
  \bibinfo{author}{\bibfnamefont{M.~D.} \bibnamefont{Coutinho-Neto}},
  \bibinfo{author}{\bibfnamefont{I.}~\bibnamefont{Tavernelli}},
  \bibnamefont{and}
  \bibinfo{author}{\bibfnamefont{U.}~\bibnamefont{Rothlisberger}},
  \bibinfo{journal}{J. Phys. Chem. B} \textbf{\bibinfo{volume}{113}},
  \bibinfo{pages}{1127} (\bibinfo{year}{2009}).

\bibitem[{\citenamefont{Schmidt et~al.}(2009)\citenamefont{Schmidt,
  VandeVondele, Kuo, Sebastiani, Siepmann, Hutter, and
  Mundy}}]{schmidt_jpcB_09}
\bibinfo{author}{\bibfnamefont{J.}~\bibnamefont{Schmidt}},
  \bibinfo{author}{\bibfnamefont{J.}~\bibnamefont{VandeVondele}},
  \bibinfo{author}{\bibfnamefont{I.-F.~W.} \bibnamefont{Kuo}},
  \bibinfo{author}{\bibfnamefont{D.}~\bibnamefont{Sebastiani}},
  \bibinfo{author}{\bibfnamefont{J.~I.} \bibnamefont{Siepmann}},
  \bibinfo{author}{\bibfnamefont{J.}~\bibnamefont{Hutter}}, \bibnamefont{and}
  \bibinfo{author}{\bibfnamefont{C.~J.} \bibnamefont{Mundy}},
  \bibinfo{journal}{J. Phys. Chem. B} \textbf{\bibinfo{volume}{113}},
  \bibinfo{pages}{11959} (\bibinfo{year}{2009}).

\bibitem[{\citenamefont{Wang et~al.}(2011)\citenamefont{Wang,
  Rom\'{a}n-P\'{e}rez, Soler, Artacho, and
  Fern\'{a}ndez-Serra}}]{wang_jcp_2011}
\bibinfo{author}{\bibfnamefont{J.}~\bibnamefont{Wang}},
  \bibinfo{author}{\bibfnamefont{G.}~\bibnamefont{Rom\'{a}n-P\'{e}rez}},
  \bibinfo{author}{\bibfnamefont{J.~M.} \bibnamefont{Soler}},
  \bibinfo{author}{\bibfnamefont{E.}~\bibnamefont{Artacho}}, \bibnamefont{and}
  \bibinfo{author}{\bibfnamefont{M.-V.} \bibnamefont{Fern\'{a}ndez-Serra}},
  \bibinfo{journal}{J. Chem. Phys.} \textbf{\bibinfo{volume}{134}},
  \bibinfo{pages}{024516} (\bibinfo{year}{2011}).

\bibitem[{\citenamefont{M{\o}gelh{\o}j
  et~al.}(2011)\citenamefont{M{\o}gelh{\o}j, Kelkkanen, Wikfeldt, Schi{\o}tz,
  Mortensen, Pettersson, Lundqvist, Jacobsen, Nilsson, and
  N{\o}rskov}}]{mogelhoj_jpcb_2011}
\bibinfo{author}{\bibfnamefont{A.}~\bibnamefont{M{\o}gelh{\o}j}},
  \bibinfo{author}{\bibfnamefont{A.~K.} \bibnamefont{Kelkkanen}},
  \bibinfo{author}{\bibfnamefont{K.~T.} \bibnamefont{Wikfeldt}},
  \bibinfo{author}{\bibfnamefont{J.}~\bibnamefont{Schi{\o}tz}},
  \bibinfo{author}{\bibfnamefont{J.~J.} \bibnamefont{Mortensen}},
  \bibinfo{author}{\bibfnamefont{L.~G.~M.} \bibnamefont{Pettersson}},
  \bibinfo{author}{\bibfnamefont{B.~I.} \bibnamefont{Lundqvist}},
  \bibinfo{author}{\bibfnamefont{K.~W.} \bibnamefont{Jacobsen}},
  \bibinfo{author}{\bibfnamefont{A.}~\bibnamefont{Nilsson}}, \bibnamefont{and}
  \bibinfo{author}{\bibfnamefont{J.~K.} \bibnamefont{N{\o}rskov}},
  \bibinfo{journal}{J. Phys. Chem. B} \textbf{\bibinfo{volume}{115}},
  \bibinfo{pages}{14149} (\bibinfo{year}{2011}).

\bibitem[{\citenamefont{Zhang et~al.}(2011{\natexlab{b}})\citenamefont{Zhang,
  Wu, Galli, and Gygi}}]{zhang_jctc_2011}
\bibinfo{author}{\bibfnamefont{C.}~\bibnamefont{Zhang}},
  \bibinfo{author}{\bibfnamefont{J.}~\bibnamefont{Wu}},
  \bibinfo{author}{\bibfnamefont{G.}~\bibnamefont{Galli}}, \bibnamefont{and}
  \bibinfo{author}{\bibfnamefont{F.}~\bibnamefont{Gygi}}, \bibinfo{journal}{J.
  Chem. Theory Comput.} \textbf{\bibinfo{volume}{7}}, \bibinfo{pages}{3054}
  (\bibinfo{year}{2011}{\natexlab{b}}).

\bibitem[{\citenamefont{Jonchiere et~al.}(2011)\citenamefont{Jonchiere,
  Seitsonen, Ferlat, Saitta, and Vuilleumier}}]{jonchiere_jcp_2011}
\bibinfo{author}{\bibfnamefont{R.}~\bibnamefont{Jonchiere}},
  \bibinfo{author}{\bibfnamefont{A.~P.} \bibnamefont{Seitsonen}},
  \bibinfo{author}{\bibfnamefont{G.}~\bibnamefont{Ferlat}},
  \bibinfo{author}{\bibfnamefont{A.~M.} \bibnamefont{Saitta}},
  \bibnamefont{and}
  \bibinfo{author}{\bibfnamefont{R.}~\bibnamefont{Vuilleumier}},
  \bibinfo{journal}{J. Chem. Phys.} \textbf{\bibinfo{volume}{135}},
  \bibinfo{pages}{154503} (\bibinfo{year}{2011}).

\bibitem[{\citenamefont{Yoo and Xantheas}(2012)}]{yoo_jcp_11}
\bibinfo{author}{\bibfnamefont{S.}~\bibnamefont{Yoo}} \bibnamefont{and}
  \bibinfo{author}{\bibfnamefont{S.~S.} \bibnamefont{Xantheas}},
  \bibinfo{journal}{J. Chem. Phys.} \textbf{\bibinfo{volume}{134}},
  \bibinfo{pages}{121105} (\bibinfo{year}{2012}).

\bibitem[{\citenamefont{Ma et~al.}(2012)\citenamefont{Ma, Zhang, and
  Tuckerman}}]{tuckerman_jcp_2012}
\bibinfo{author}{\bibfnamefont{Z.}~\bibnamefont{Ma}},
  \bibinfo{author}{\bibfnamefont{Y.}~\bibnamefont{Zhang}}, \bibnamefont{and}
  \bibinfo{author}{\bibfnamefont{M.~E.} \bibnamefont{Tuckerman}},
  \bibinfo{journal}{J. Chem. Phys.} \textbf{\bibinfo{volume}{137}},
  \bibinfo{pages}{044506} (\bibinfo{year}{2012}).

\bibitem[{zha()}]{zhaofeng_thesis_2012}
\bibinfo{note}{Z. Li, \emph{Improving Ab Initio Molecular Dynamics of Liquid
  Water}, Ph.D. Thesis, Princeton University, 2012},
  \url{http://www.princeton.edu/physics/graduate-program/theses/Thes%
is_Zhaofeng-Li.pdf}.

\bibitem[{\citenamefont{Dion et~al.}(2004)\citenamefont{Dion, Rydberg,
  Schr\"{o}der, Langreth, and Lundqvist}}]{Dion_vdw_04}
\bibinfo{author}{\bibfnamefont{M.}~\bibnamefont{Dion}},
  \bibinfo{author}{\bibfnamefont{H.}~\bibnamefont{Rydberg}},
  \bibinfo{author}{\bibfnamefont{E.}~\bibnamefont{Schr\"{o}der}},
  \bibinfo{author}{\bibfnamefont{D.~C.} \bibnamefont{Langreth}},
  \bibnamefont{and} \bibinfo{author}{\bibfnamefont{B.~I.}
  \bibnamefont{Lundqvist}}, \bibinfo{journal}{Phys. Rev. Lett.}
  \textbf{\bibinfo{volume}{92}}, \bibinfo{pages}{246401}
  (\bibinfo{year}{2004}).

\bibitem[{\citenamefont{Tkatchenko and Scheffler}(2009)}]{TS_vdw_09}
\bibinfo{author}{\bibfnamefont{A.}~\bibnamefont{Tkatchenko}} \bibnamefont{and}
  \bibinfo{author}{\bibfnamefont{M.}~\bibnamefont{Scheffler}},
  \bibinfo{journal}{Phys. Rev. Lett.} \textbf{\bibinfo{volume}{102}},
  \bibinfo{pages}{073005} (\bibinfo{year}{2009}).

\bibitem[{\citenamefont{Klime\v{s} et~al.}(2010)\citenamefont{Klime\v{s},
  Bowler, and Michaelides}}]{klimes_10}
\bibinfo{author}{\bibfnamefont{J.}~\bibnamefont{Klime\v{s}}},
  \bibinfo{author}{\bibfnamefont{D.~R.} \bibnamefont{Bowler}},
  \bibnamefont{and}
  \bibinfo{author}{\bibfnamefont{A.}~\bibnamefont{Michaelides}},
  \bibinfo{journal}{J. Phys.: Cond. Matt.} \textbf{\bibinfo{volume}{22}},
  \bibinfo{pages}{022201} (\bibinfo{year}{2010}).

\bibitem[{\citenamefont{von Lilienfeld et~al.}(2004)\citenamefont{von
  Lilienfeld, Tavernelli, Rothlisberger, and Sebastiani}}]{oavl_prl_2004}
\bibinfo{author}{\bibfnamefont{O.~A.} \bibnamefont{von Lilienfeld}},
  \bibinfo{author}{\bibfnamefont{I.}~\bibnamefont{Tavernelli}},
  \bibinfo{author}{\bibfnamefont{U.}~\bibnamefont{Rothlisberger}},
  \bibnamefont{and}
  \bibinfo{author}{\bibfnamefont{D.}~\bibnamefont{Sebastiani}},
  \bibinfo{journal}{Phys. Rev. Lett.} \textbf{\bibinfo{volume}{93}},
  \bibinfo{pages}{153004} (\bibinfo{year}{2004}).

\bibitem[{\citenamefont{Silvestrelli}(2008)}]{silvestrelli_prl_2008}
\bibinfo{author}{\bibfnamefont{P.~L.} \bibnamefont{Silvestrelli}},
  \bibinfo{journal}{Phys. Rev. Lett.} \textbf{\bibinfo{volume}{100}},
  \bibinfo{pages}{053002} (\bibinfo{year}{2008}).

\bibitem[{\citenamefont{Sato and Nakai}(2009)}]{sato_jcp_2009}
\bibinfo{author}{\bibfnamefont{T.}~\bibnamefont{Sato}} \bibnamefont{and}
  \bibinfo{author}{\bibfnamefont{H.}~\bibnamefont{Nakai}}, \bibinfo{journal}{J.
  Chem. Phys.} \textbf{\bibinfo{volume}{131}}, \bibinfo{pages}{224104}
  (\bibinfo{year}{2009}).

\bibitem[{\citenamefont{Grimme et~al.}(2010)\citenamefont{Grimme, Antony,
  Ehrlich, and Krieg}}]{grimme_vdw_10}
\bibinfo{author}{\bibfnamefont{S.}~\bibnamefont{Grimme}},
  \bibinfo{author}{\bibfnamefont{J.}~\bibnamefont{Antony}},
  \bibinfo{author}{\bibfnamefont{S.}~\bibnamefont{Ehrlich}}, \bibnamefont{and}
  \bibinfo{author}{\bibfnamefont{H.}~\bibnamefont{Krieg}}, \bibinfo{journal}{J.
  Chem. Phys.} \textbf{\bibinfo{volume}{132}}, \bibinfo{pages}{154104}
  (\bibinfo{year}{2010}).

\bibitem[{\citenamefont{Vydrov and Voorhis}(2009)}]{vydrov_prl_2009}
\bibinfo{author}{\bibfnamefont{O.~A.} \bibnamefont{Vydrov}} \bibnamefont{and}
  \bibinfo{author}{\bibfnamefont{T.~V.} \bibnamefont{Voorhis}},
  \bibinfo{journal}{Phys. Rev. Lett.} \textbf{\bibinfo{volume}{103}},
  \bibinfo{pages}{063004} (\bibinfo{year}{2009}).

\bibitem[{\citenamefont{Becke and Johnson}(2005)}]{becke_jcp_2005}
\bibinfo{author}{\bibfnamefont{A.~D.} \bibnamefont{Becke}} \bibnamefont{and}
  \bibinfo{author}{\bibfnamefont{E.~R.} \bibnamefont{Johnson}},
  \bibinfo{journal}{J. Chem. Phys.} \textbf{\bibinfo{volume}{123}},
  \bibinfo{pages}{154101} (\bibinfo{year}{2005}).

\bibitem[{\citenamefont{Tao et~al.}(2012)\citenamefont{Tao, Perdew, and
  Ruzsinszky}}]{perdew_pnas_2012}
\bibinfo{author}{\bibfnamefont{J.}~\bibnamefont{Tao}},
  \bibinfo{author}{\bibfnamefont{J.~P.} \bibnamefont{Perdew}},
  \bibnamefont{and}
  \bibinfo{author}{\bibfnamefont{A.}~\bibnamefont{Ruzsinszky}},
  \bibinfo{journal}{Proc. Natl. Acad. Sci. (USA)}
  \textbf{\bibinfo{volume}{109}}, \bibinfo{pages}{18} (\bibinfo{year}{2012}).

\bibitem[{\citenamefont{Petrenko and Whitworth}(2003)}]{ice_book}
\bibinfo{author}{\bibfnamefont{V.}~\bibnamefont{Petrenko}} \bibnamefont{and}
  \bibinfo{author}{\bibfnamefont{R.~W.} \bibnamefont{Whitworth}},
  \emph{\bibinfo{title}{Physics of Ice}} (\bibinfo{publisher}{Oxford University
  Press}, \bibinfo{address}{New York}, \bibinfo{year}{2003}).

\bibitem[{\citenamefont{Salzmann et~al.}(2009)\citenamefont{Salzmann, Radaelli,
  Mayer, and Finney}}]{ice-15}
\bibinfo{author}{\bibfnamefont{C.~G.} \bibnamefont{Salzmann}},
  \bibinfo{author}{\bibfnamefont{P.~G.} \bibnamefont{Radaelli}},
  \bibinfo{author}{\bibfnamefont{E.}~\bibnamefont{Mayer}}, \bibnamefont{and}
  \bibinfo{author}{\bibfnamefont{J.~L.} \bibnamefont{Finney}},
  \bibinfo{journal}{Phys. Rev. Lett.} \textbf{\bibinfo{volume}{103}},
  \bibinfo{pages}{105701} (\bibinfo{year}{2009}).

\bibitem[{sal()}]{salzmann_rsc_2007}
\bibinfo{note}{C. G. Salzmann, P. G. Radaelli, A. Hallbrucker, E. Mayer, and J.
  L. Finney, in{ \it Physics and Chemistry of Ice}, ed. W. F. Kuhs, The Royal
  Society of Chemistry, Cambridge, 2007, pp. 521–528}.

\bibitem[{\citenamefont{Salzmann et~al.}(2011)\citenamefont{Salzmann, Radaelli,
  Slater, and Finney}}]{salzmann_pccp_2011}
\bibinfo{author}{\bibfnamefont{C.~G.} \bibnamefont{Salzmann}},
  \bibinfo{author}{\bibfnamefont{P.~G.} \bibnamefont{Radaelli}},
  \bibinfo{author}{\bibfnamefont{B.}~\bibnamefont{Slater}}, \bibnamefont{and}
  \bibinfo{author}{\bibfnamefont{J.~L.} \bibnamefont{Finney}},
  \bibinfo{journal}{Phys. Chem. Chem. Phys.} \textbf{\bibinfo{volume}{13}},
  \bibinfo{pages}{18468} (\bibinfo{year}{2011}).

\bibitem[{\citenamefont{Tkatchenko et~al.}(2012)\citenamefont{Tkatchenko,
  DiStasio~Jr., Car, and Scheffler}}]{MBD_prl_12}
\bibinfo{author}{\bibfnamefont{A.}~\bibnamefont{Tkatchenko}},
  \bibinfo{author}{\bibfnamefont{R.~A.} \bibnamefont{DiStasio~Jr.}},
  \bibinfo{author}{\bibfnamefont{R.}~\bibnamefont{Car}}, \bibnamefont{and}
  \bibinfo{author}{\bibfnamefont{M.}~\bibnamefont{Scheffler}},
  \bibinfo{journal}{Phys. Rev. Lett.} \textbf{\bibinfo{volume}{108}},
  \bibinfo{pages}{236402} (\bibinfo{year}{2012}).

\bibitem[{\citenamefont{Lee et~al.}(2010)\citenamefont{Lee, Murray, Kong,
  Lundqvist, and Langreth}}]{vdW-DF2}
\bibinfo{author}{\bibfnamefont{K.}~\bibnamefont{Lee}},
  \bibinfo{author}{\bibfnamefont{E.~D.} \bibnamefont{Murray}},
  \bibinfo{author}{\bibfnamefont{L.}~\bibnamefont{Kong}},
  \bibinfo{author}{\bibfnamefont{B.~I.} \bibnamefont{Lundqvist}},
  \bibnamefont{and} \bibinfo{author}{\bibfnamefont{D.~C.}
  \bibnamefont{Langreth}}, \bibinfo{journal}{Phys. Rev. B}
  \textbf{\bibinfo{volume}{82}}, \bibinfo{pages}{081101(R)}
  (\bibinfo{year}{2010}).

\bibitem[{\citenamefont{Langreth{ \it et al.}}(2009)}]{langreth_2009}
\bibinfo{author}{\bibfnamefont{D.~C.} \bibnamefont{Langreth{ \it et al.}}},
  \bibinfo{journal}{J. Phys.: Condens. Matter} \textbf{\bibinfo{volume}{21}},
  \bibinfo{pages}{084203} (\bibinfo{year}{2009}).

\bibitem[{\citenamefont{Tkatchenko et~al.}(2010)\citenamefont{Tkatchenko,
  Romaner, Hofmann, Zojer, and Ambrosch-Draxl}}]{tkatchenko_mrs_2010}
\bibinfo{author}{\bibfnamefont{A.}~\bibnamefont{Tkatchenko}},
  \bibinfo{author}{\bibfnamefont{L.}~\bibnamefont{Romaner}},
  \bibinfo{author}{\bibfnamefont{O.~T.} \bibnamefont{Hofmann}},
  \bibinfo{author}{\bibfnamefont{E.}~\bibnamefont{Zojer}}, \bibnamefont{and}
  \bibinfo{author}{\bibfnamefont{C.}~\bibnamefont{Ambrosch-Draxl}},
  \bibinfo{journal}{MRS Bull.} \textbf{\bibinfo{volume}{35}},
  \bibinfo{pages}{435} (\bibinfo{year}{2010}).

\bibitem[{\citenamefont{Klime\v{s} et~al.}(2011)\citenamefont{Klime\v{s},
  Bowler, and Michaelides}}]{klimes_PRB_11}
\bibinfo{author}{\bibfnamefont{J.}~\bibnamefont{Klime\v{s}}},
  \bibinfo{author}{\bibfnamefont{D.~R.} \bibnamefont{Bowler}},
  \bibnamefont{and}
  \bibinfo{author}{\bibfnamefont{A.}~\bibnamefont{Michaelides}},
  \bibinfo{journal}{Phys. Rev. B} \textbf{\bibinfo{volume}{83}},
  \bibinfo{pages}{195131} (\bibinfo{year}{2011}).

\bibitem[{\citenamefont{{DiStasio Jr.} et~al.}(2012)\citenamefont{{DiStasio
  Jr.}, {von Lilienfeld}, and Tkatchenko}}]{distasio_pnas_12}
\bibinfo{author}{\bibfnamefont{R.~A.} \bibnamefont{{DiStasio Jr.}}},
  \bibinfo{author}{\bibfnamefont{O.~A.} \bibnamefont{{von Lilienfeld}}},
  \bibnamefont{and}
  \bibinfo{author}{\bibfnamefont{A.}~\bibnamefont{Tkatchenko}},
  \bibinfo{journal}{Proc. Natl. Acad. Sci. (USA)}
  \textbf{\bibinfo{volume}{109}}, \bibinfo{pages}{14791}
  (\bibinfo{year}{2012}).

\bibitem[{\citenamefont{Klime\v{s} and Michaelides}(2012)}]{klimes_jcp_2012}
\bibinfo{author}{\bibfnamefont{J.}~\bibnamefont{Klime\v{s}}} \bibnamefont{and}
  \bibinfo{author}{\bibfnamefont{A.}~\bibnamefont{Michaelides}},
  \bibinfo{journal}{J. Chem. Phys.} \textbf{\bibinfo{volume}{137}},
  \bibinfo{pages}{120901} (\bibinfo{year}{2012}).

\bibitem[{\citenamefont{Reilly and Tkatchenko}(2013)}]{reilly_jpcl_2013}
\bibinfo{author}{\bibfnamefont{A.~M.} \bibnamefont{Reilly}} \bibnamefont{and}
  \bibinfo{author}{\bibfnamefont{A.}~\bibnamefont{Tkatchenko}},
  \bibinfo{journal}{J. Phys. Chem. Lett.} \textbf{\bibinfo{volume}{4}},
  \bibinfo{pages}{1028} (\bibinfo{year}{2013}).

\bibitem[{\citenamefont{Kelkkanen et~al.}(2009)\citenamefont{Kelkkanen,
  Lundqvist, and N{\o}rskov}}]{kelkkanen_jcp_2009}
\bibinfo{author}{\bibfnamefont{A.~K.} \bibnamefont{Kelkkanen}},
  \bibinfo{author}{\bibfnamefont{B.~I.} \bibnamefont{Lundqvist}},
  \bibnamefont{and} \bibinfo{author}{\bibfnamefont{J.~K.}
  \bibnamefont{N{\o}rskov}}, \bibinfo{journal}{J. Chem. Phys.}
  \textbf{\bibinfo{volume}{131}}, \bibinfo{pages}{046102}
  (\bibinfo{year}{2009}).

\bibitem[{\citenamefont{Line and Whitworth}(1996)}]{ice-XI_96}
\bibinfo{author}{\bibfnamefont{C.~M.~B.} \bibnamefont{Line}} \bibnamefont{and}
  \bibinfo{author}{\bibfnamefont{R.~W.} \bibnamefont{Whitworth}},
  \bibinfo{journal}{J. Chem. Phys.} \textbf{\bibinfo{volume}{104}},
  \bibinfo{pages}{10008} (\bibinfo{year}{1996}).

\bibitem[{\citenamefont{Londono et~al.}(1993)\citenamefont{Londono, Kuhs, and
  Finney}}]{ice-9_93}
\bibinfo{author}{\bibfnamefont{J.~D.} \bibnamefont{Londono}},
  \bibinfo{author}{\bibfnamefont{W.~F.} \bibnamefont{Kuhs}}, \bibnamefont{and}
  \bibinfo{author}{\bibfnamefont{J.~L.} \bibnamefont{Finney}},
  \bibinfo{journal}{J. Chem. Phys.} \textbf{\bibinfo{volume}{98}},
  \bibinfo{pages}{4878} (\bibinfo{year}{1993}).

\bibitem[{\citenamefont{Lobban et~al.}(2002)\citenamefont{Lobban, Finney, and
  Kuhs}}]{ice-2}
\bibinfo{author}{\bibfnamefont{C.}~\bibnamefont{Lobban}},
  \bibinfo{author}{\bibfnamefont{J.~L.} \bibnamefont{Finney}},
  \bibnamefont{and} \bibinfo{author}{\bibfnamefont{W.~F.} \bibnamefont{Kuhs}},
  \bibinfo{journal}{J. Chem. Phys.} \textbf{\bibinfo{volume}{117}},
  \bibinfo{pages}{3928} (\bibinfo{year}{2002}).

\bibitem[{\citenamefont{Salzmann et~al.}(2006)\citenamefont{Salzmann, Radaelli,
  Hallbrucker, Mayer, and Finney}}]{ice-13-14}
\bibinfo{author}{\bibfnamefont{C.~G.} \bibnamefont{Salzmann}},
  \bibinfo{author}{\bibfnamefont{P.~G.} \bibnamefont{Radaelli}},
  \bibinfo{author}{\bibfnamefont{A.}~\bibnamefont{Hallbrucker}},
  \bibinfo{author}{\bibfnamefont{E.}~\bibnamefont{Mayer}}, \bibnamefont{and}
  \bibinfo{author}{\bibfnamefont{J.~L.} \bibnamefont{Finney}},
  \bibinfo{journal}{Science} \textbf{\bibinfo{volume}{311}},
  \bibinfo{pages}{1758} (\bibinfo{year}{2006}).

\bibitem[{\citenamefont{Kuhs et~al.}(1984)\citenamefont{Kuhs, Finney, Vettier,
  and Bliss}}]{ice-8}
\bibinfo{author}{\bibfnamefont{W.~F.} \bibnamefont{Kuhs}},
  \bibinfo{author}{\bibfnamefont{J.~L.} \bibnamefont{Finney}},
  \bibinfo{author}{\bibfnamefont{C.}~\bibnamefont{Vettier}}, \bibnamefont{and}
  \bibinfo{author}{\bibfnamefont{D.~V.} \bibnamefont{Bliss}},
  \bibinfo{journal}{J. Chem. Phys.} \textbf{\bibinfo{volume}{81}},
  \bibinfo{pages}{3612} (\bibinfo{year}{1984}).

\bibitem[{\citenamefont{Pan et~al.}(2010)\citenamefont{Pan, Liu, Tribello,
  Slater, Michaelides, and Wang}}]{pan_10}
\bibinfo{author}{\bibfnamefont{D.}~\bibnamefont{Pan}},
  \bibinfo{author}{\bibfnamefont{L.-M.} \bibnamefont{Liu}},
  \bibinfo{author}{\bibfnamefont{G.~A.} \bibnamefont{Tribello}},
  \bibinfo{author}{\bibfnamefont{B.}~\bibnamefont{Slater}},
  \bibinfo{author}{\bibfnamefont{A.}~\bibnamefont{Michaelides}},
  \bibnamefont{and} \bibinfo{author}{\bibfnamefont{E.}~\bibnamefont{Wang}},
  \bibinfo{journal}{J. Phys.: Condensed Matter 22}
  \textbf{\bibinfo{volume}{22}}, \bibinfo{pages}{074203}
  (\bibinfo{year}{2010}).

\bibitem[{\citenamefont{Perdew et~al.}(1996)\citenamefont{Perdew, Burke, and
  Ernzerhof}}]{PBE}
\bibinfo{author}{\bibfnamefont{J.~P.} \bibnamefont{Perdew}},
  \bibinfo{author}{\bibfnamefont{K.}~\bibnamefont{Burke}}, \bibnamefont{and}
  \bibinfo{author}{\bibfnamefont{M.}~\bibnamefont{Ernzerhof}},
  \bibinfo{journal}{Phys. Rev. Lett.} \textbf{\bibinfo{volume}{77}},
  \bibinfo{pages}{3865} (\bibinfo{year}{1996}).

\bibitem[{mur()}]{murnaghan44}
\bibinfo{note}{F. D. Murnaghan, Proc. Natl. Acad. Sci. USA {\bf 30}, 244
  (1944).}

\bibitem[{san()}]{santra_thesis_2010}
\bibinfo{note}{B. Santra, \emph{Density-Functional Theory Exchange-Correlation
  Functionals for Hydrogen Bonds in Water}, Ph.D. Thesis, Fritz-Haber-Institut
  der Max-Planck Gesellschaft, TU-Berlin, 2010},
  \url{http://opus4.kobv.de/opus4-tuberlin/frontdoor/index/index/doc%
Id/2723}.

\bibitem[{\citenamefont{Adamo and Barone}(1999)}]{PBE0}
\bibinfo{author}{\bibfnamefont{C.}~\bibnamefont{Adamo}} \bibnamefont{and}
  \bibinfo{author}{\bibfnamefont{V.}~\bibnamefont{Barone}},
  \bibinfo{journal}{J. Chem. Phys.} \textbf{\bibinfo{volume}{110}},
  \bibinfo{pages}{6158} (\bibinfo{year}{1999}).

\bibitem[{\citenamefont{Zhang and Yang}(1998)}]{revPBE}
\bibinfo{author}{\bibfnamefont{Y.}~\bibnamefont{Zhang}} \bibnamefont{and}
  \bibinfo{author}{\bibfnamefont{W.}~\bibnamefont{Yang}},
  \bibinfo{journal}{Phys. Rev. Lett.} \textbf{\bibinfo{volume}{80}},
  \bibinfo{pages}{890} (\bibinfo{year}{1998}).

\bibitem[{\citenamefont{Jure\v{c}ka et~al.}(2006)\citenamefont{Jure\v{c}ka,
  \v{S}poner, \v{C}ern\'{y}, and Hobza}}]{s22}
\bibinfo{author}{\bibfnamefont{P.}~\bibnamefont{Jure\v{c}ka}},
  \bibinfo{author}{\bibfnamefont{J.}~\bibnamefont{\v{S}poner}},
  \bibinfo{author}{\bibfnamefont{J.}~\bibnamefont{\v{C}ern\'{y}}},
  \bibnamefont{and} \bibinfo{author}{\bibfnamefont{P.}~\bibnamefont{Hobza}},
  \bibinfo{journal}{Phys. Chem. Chem. Phys.} \textbf{\bibinfo{volume}{8}},
  \bibinfo{pages}{1985} (\bibinfo{year}{2006}).

\bibitem[{\citenamefont{Murray et~al.}(2009)\citenamefont{Murray, Lee, and
  Langreth}}]{rPW86_2009}
\bibinfo{author}{\bibfnamefont{E.~D.} \bibnamefont{Murray}},
  \bibinfo{author}{\bibfnamefont{K.}~\bibnamefont{Lee}}, \bibnamefont{and}
  \bibinfo{author}{\bibfnamefont{D.~C.} \bibnamefont{Langreth}},
  \bibinfo{journal}{J. Chem. Theory Comput.} \textbf{\bibinfo{volume}{5}},
  \bibinfo{pages}{2754} (\bibinfo{year}{2009}).

\bibitem[{\citenamefont{Blum{ \it et al.}}(2009)}]{FHI-aims}
\bibinfo{author}{\bibfnamefont{V.}~\bibnamefont{Blum{ \it et al.}}},
  \bibinfo{journal}{Comp. Phys. Comm.} \textbf{\bibinfo{volume}{180}},
  \bibinfo{pages}{2175} (\bibinfo{year}{2009}).

\bibitem[{\citenamefont{Kresse and Hafner}(1993)}]{vasp-1}
\bibinfo{author}{\bibfnamefont{G.}~\bibnamefont{Kresse}} \bibnamefont{and}
  \bibinfo{author}{\bibfnamefont{J.}~\bibnamefont{Hafner}},
  \bibinfo{journal}{Phys. Rev. B} \textbf{\bibinfo{volume}{47}},
  \bibinfo{pages}{558} (\bibinfo{year}{1993}).

\bibitem[{\citenamefont{Kresse and Furthm\"uller}(1996)}]{vasp-2}
\bibinfo{author}{\bibfnamefont{G.}~\bibnamefont{Kresse}} \bibnamefont{and}
  \bibinfo{author}{\bibfnamefont{J.}~\bibnamefont{Furthm\"uller}},
  \bibinfo{journal}{Phys. Rev. B} \textbf{\bibinfo{volume}{54}},
  \bibinfo{pages}{11169} (\bibinfo{year}{1996}).

\bibitem[{\citenamefont{Rom\'{a}n-P\'{e}rez and Soler}(2009)}]{soler_09}
\bibinfo{author}{\bibfnamefont{G.}~\bibnamefont{Rom\'{a}n-P\'{e}rez}}
  \bibnamefont{and} \bibinfo{author}{\bibfnamefont{J.~M.} \bibnamefont{Soler}},
  \bibinfo{journal}{Phys. Rev. Lett.} \textbf{\bibinfo{volume}{103}},
  \bibinfo{pages}{096102} (\bibinfo{year}{2009}).

\bibitem[{\citenamefont{Whalley}(1984)}]{whalley_jcp_84}
\bibinfo{author}{\bibfnamefont{E.}~\bibnamefont{Whalley}}, \bibinfo{journal}{J.
  Chem. Phys.} \textbf{\bibinfo{volume}{81}}, \bibinfo{pages}{4087}
  (\bibinfo{year}{1984}).

\bibitem[{\citenamefont{R\"ottger et~al.}(1994)\citenamefont{R\"ottger,
  Endriss, Ihringer, Doyle, and Kuhs}}]{rottger_Ih_94}
\bibinfo{author}{\bibfnamefont{K.}~\bibnamefont{R\"ottger}},
  \bibinfo{author}{\bibfnamefont{A.}~\bibnamefont{Endriss}},
  \bibinfo{author}{\bibfnamefont{J.}~\bibnamefont{Ihringer}},
  \bibinfo{author}{\bibfnamefont{S.}~\bibnamefont{Doyle}}, \bibnamefont{and}
  \bibinfo{author}{\bibfnamefont{W.~F.} \bibnamefont{Kuhs}},
  \bibinfo{journal}{Acta Cryst.} \textbf{\bibinfo{volume}{B50}},
  \bibinfo{pages}{644} (\bibinfo{year}{1994}).

\bibitem[{\citenamefont{Carrasco et~al.}(2011)\citenamefont{Carrasco, Santra,
  Klime\v{s}, and Michaelides}}]{carrasco_prl_2011}
\bibinfo{author}{\bibfnamefont{J.}~\bibnamefont{Carrasco}},
  \bibinfo{author}{\bibfnamefont{B.}~\bibnamefont{Santra}},
  \bibinfo{author}{\bibfnamefont{J.}~\bibnamefont{Klime\v{s}}},
  \bibnamefont{and}
  \bibinfo{author}{\bibfnamefont{A.}~\bibnamefont{Michaelides}},
  \bibinfo{journal}{Phys. Rev. Lett.} \textbf{\bibinfo{volume}{106}},
  \bibinfo{pages}{026101} (\bibinfo{year}{2011}).

\bibitem[{\citenamefont{Gulans et~al.}(2009)\citenamefont{Gulans, Puska, and
  Nieminen}}]{gulans_prb_2009}
\bibinfo{author}{\bibfnamefont{A.}~\bibnamefont{Gulans}},
  \bibinfo{author}{\bibfnamefont{M.~J.} \bibnamefont{Puska}}, \bibnamefont{and}
  \bibinfo{author}{\bibfnamefont{R.~M.} \bibnamefont{Nieminen}},
  \bibinfo{journal}{Phys. Rev. B} \textbf{\bibinfo{volume}{79}},
  \bibinfo{pages}{201105(R)} (\bibinfo{year}{2009}).

\bibitem[{not({\natexlab{a}})}]{note_on_PBE0+vdW}
\bibinfo{note}{Frequency calculations have not been performed for any of the
  PBE0 based functionals. For PBE0 we have estimated the ZPE corrected volumes
  by considering the volume changes found with PBE. For PBE0+vdW$^{\rm TS}$ and
  PBE0+vdW$^{\rm MB}$ ZPE corrected volumes are estimated with the volume
  changes obtained with PBE+vdW$^{\rm TS}$.}

\bibitem[{\citenamefont{Kell and Whalley}(1968)}]{kell_jcp_1968}
\bibinfo{author}{\bibfnamefont{G.~S.} \bibnamefont{Kell}} \bibnamefont{and}
  \bibinfo{author}{\bibfnamefont{E.}~\bibnamefont{Whalley}},
  \bibinfo{journal}{J. Chem. Phys.} \textbf{\bibinfo{volume}{48}},
  \bibinfo{pages}{2359} (\bibinfo{year}{1968}).

\bibitem[{not({\natexlab{b}})}]{note_on_c6_calculation}
\bibinfo{note}{In case of vdW$^{\rm TS}$ molecular $C_6$ coefficients are
  calculated by using Eq. 10 in Ref.~\cite{TS_vdw_09} and for the vdW-DFs we
  have used the long distance behavior of the kernel as given by Eq. 17 in
  Ref.~\cite{Dion_vdw_04} and in Eq. 23.5 in Ref.~\cite{vydrov2011chap}. The
  vdW-DF $C_6$ coefficients vary only within 1\% when using different PAW
  potentials or norm-conserving pseudopotentials, however, they depend
  significantly on the functional employed to calculate the electron density.
  For example, the $C_6$ coefficient of an isolated water molecule is reduced
  to 45.5 a.u. when Hartree-Fock electron density is used. Likewise, the vdW-DF
  $C_6$ coefficients of an isolated water molecule obtained here are 20-24\%
  larger than the values reported in Ref.~\cite{vydrov_pra_2010} where electron
  densities were calculated using a so-called long-range corrected hybrid
  functional which is expected to give densities similar to those given by
  Hartree-Fock.}

\bibitem[{\citenamefont{Vydrov and Voorhis}(2010)}]{vydrov_pra_2010}
\bibinfo{author}{\bibfnamefont{O.~A.} \bibnamefont{Vydrov}} \bibnamefont{and}
  \bibinfo{author}{\bibfnamefont{T.~V.} \bibnamefont{Voorhis}},
  \bibinfo{journal}{Phys. Rev. A} \textbf{\bibinfo{volume}{81}},
  \bibinfo{pages}{062708} (\bibinfo{year}{2010}).

\bibitem[{\citenamefont{Burns et~al.}(2011)\citenamefont{Burns, \'{A}lvaro
  V\'{a}zquez-Mayagoitia, Sumpter, and Sherrill}}]{burns_jcp_11}
\bibinfo{author}{\bibfnamefont{L.~A.} \bibnamefont{Burns}},
  \bibinfo{author}{\bibnamefont{\'{A}lvaro V\'{a}zquez-Mayagoitia}},
  \bibinfo{author}{\bibfnamefont{B.~G.} \bibnamefont{Sumpter}},
  \bibnamefont{and} \bibinfo{author}{\bibfnamefont{C.~D.}
  \bibnamefont{Sherrill}}, \bibinfo{journal}{J. Chem. Phys.}
  \textbf{\bibinfo{volume}{134}}, \bibinfo{pages}{084107}
  (\bibinfo{year}{2011}).

\bibitem[{\citenamefont{Lu et~al.}(2009)\citenamefont{Lu, Li, Rocca, and
  Galli}}]{lu_prl_2009}
\bibinfo{author}{\bibfnamefont{D.}~\bibnamefont{Lu}},
  \bibinfo{author}{\bibfnamefont{Y.}~\bibnamefont{Li}},
  \bibinfo{author}{\bibfnamefont{D.}~\bibnamefont{Rocca}}, \bibnamefont{and}
  \bibinfo{author}{\bibfnamefont{G.}~\bibnamefont{Galli}},
  \bibinfo{journal}{Phys. Rev. Lett.} \textbf{\bibinfo{volume}{102}},
  \bibinfo{pages}{206411} (\bibinfo{year}{2009}).

\bibitem[{\citenamefont{Schimka et~al.}(2010)\citenamefont{Schimka, Harl,
  Stroppa, Gr\"uneis, Marsman, Mittendorfer, and Kresse}}]{schimka_nmat_2010}
\bibinfo{author}{\bibfnamefont{L.}~\bibnamefont{Schimka}},
  \bibinfo{author}{\bibfnamefont{J.}~\bibnamefont{Harl}},
  \bibinfo{author}{\bibfnamefont{A.}~\bibnamefont{Stroppa}},
  \bibinfo{author}{\bibfnamefont{A.}~\bibnamefont{Gr\"uneis}},
  \bibinfo{author}{\bibfnamefont{M.}~\bibnamefont{Marsman}},
  \bibinfo{author}{\bibfnamefont{F.}~\bibnamefont{Mittendorfer}},
  \bibnamefont{and} \bibinfo{author}{\bibfnamefont{G.}~\bibnamefont{Kresse}},
  \bibinfo{journal}{Nature Mater.} \textbf{\bibinfo{volume}{9}},
  \bibinfo{pages}{741} (\bibinfo{year}{2010}).

\bibitem[{\citenamefont{Gr\"uneis et~al.}(2010)\citenamefont{Gr\"uneis,
  Marsman, and Kresse}}]{gruneis_jcp_2010}
\bibinfo{author}{\bibfnamefont{A.}~\bibnamefont{Gr\"uneis}},
  \bibinfo{author}{\bibfnamefont{M.}~\bibnamefont{Marsman}}, \bibnamefont{and}
  \bibinfo{author}{\bibfnamefont{G.}~\bibnamefont{Kresse}},
  \bibinfo{journal}{J. Chem. Phys.} \textbf{\bibinfo{volume}{133}},
  \bibinfo{pages}{074107} (\bibinfo{year}{2010}).

\bibitem[{\citenamefont{Pickard and Needs}(2007)}]{pickard_2007}
\bibinfo{author}{\bibfnamefont{C.~J.} \bibnamefont{Pickard}} \bibnamefont{and}
  \bibinfo{author}{\bibfnamefont{R.~J.} \bibnamefont{Needs}},
  \bibinfo{journal}{J. Chem. Phys.} \textbf{\bibinfo{volume}{127}},
  \bibinfo{pages}{244503} (\bibinfo{year}{2007}).

\bibitem[{\citenamefont{McMahon}(2011)}]{mcmahon_prb_2011}
\bibinfo{author}{\bibfnamefont{J.~M.} \bibnamefont{McMahon}},
  \bibinfo{journal}{Phys. Rev. B} \textbf{\bibinfo{volume}{84}},
  \bibinfo{pages}{220104(R)} (\bibinfo{year}{2011}).

\bibitem[{\citenamefont{Hermanna et~al.}(2012)\citenamefont{Hermanna, Ashcroft,
  and Hoffmann}}]{hermann_pnas_2012}
\bibinfo{author}{\bibfnamefont{A.}~\bibnamefont{Hermanna}},
  \bibinfo{author}{\bibfnamefont{N.~W.} \bibnamefont{Ashcroft}},
  \bibnamefont{and} \bibinfo{author}{\bibfnamefont{R.}~\bibnamefont{Hoffmann}},
  \bibinfo{journal}{Proc. Natl. Acad. Sci. (USA)}
  \textbf{\bibinfo{volume}{109}}, \bibinfo{pages}{745} (\bibinfo{year}{2012}).

\bibitem[{\citenamefont{Pickard et~al.}(2013)\citenamefont{Pickard,
  Martinez-Canales, and Needs}}]{pickard_prl_2013}
\bibinfo{author}{\bibfnamefont{C.~J.} \bibnamefont{Pickard}},
  \bibinfo{author}{\bibfnamefont{M.}~\bibnamefont{Martinez-Canales}},
  \bibnamefont{and} \bibinfo{author}{\bibfnamefont{R.~J.} \bibnamefont{Needs}},
  \bibinfo{journal}{Phys. Rev. Lett.} \textbf{\bibinfo{volume}{110}},
  \bibinfo{pages}{245701} (\bibinfo{year}{2013}).

\bibitem[{\citenamefont{Vydrov and Van~Voorhis}(2012)}]{vydrov2011chap}
\bibinfo{author}{\bibfnamefont{O.~A.} \bibnamefont{Vydrov}} \bibnamefont{and}
  \bibinfo{author}{\bibfnamefont{T.}~\bibnamefont{Van~Voorhis}}, in
  \emph{\bibinfo{booktitle}{Fundamentals of Time-Dependent Density Functional
  Theory}}, edited by \bibinfo{editor}{\bibfnamefont{M.~A.~L.}
  \bibnamefont{Marques}},
  \bibinfo{editor}{\bibfnamefont{N.}~\bibnamefont{Maitra}},
  \bibinfo{editor}{\bibfnamefont{F.}~\bibnamefont{Nogueira}},
  \bibinfo{editor}{\bibfnamefont{E.~K.~U.} \bibnamefont{Gross}},
  \bibnamefont{and} \bibinfo{editor}{\bibfnamefont{A.}~\bibnamefont{Rubio}}
  (\bibinfo{publisher}{Springer, Berlin}, \bibinfo{year}{2012}).

\bibitem[{sup()}]{supplement}
\bibinfo{note}{Supplementary material provides coordinates of all ice phases
  optimized with all{ \it xc} functionals considered here. It can be obtained
  via \url{https://docs.google.com/file/d/0B_RP_vP8_oA5SzItT0ZSRW1Pa1U/edit}.}

\bibitem[{\citenamefont{Margoliash and Meath}(1978)}]{margoliash_jcp_1978}
\bibinfo{author}{\bibfnamefont{D.~J.} \bibnamefont{Margoliash}}
  \bibnamefont{and} \bibinfo{author}{\bibfnamefont{W.~J.} \bibnamefont{Meath}},
  \bibinfo{journal}{J. Chem. Phys.} \textbf{\bibinfo{volume}{68}},
  \bibinfo{pages}{1426} (\bibinfo{year}{1978}).

\end{thebibliography}



\end{document}